\documentclass[reprint,pre,amsmath,amssymb,aps,showkeys]{revtex4-2}

\usepackage[T1]{fontenc}
\usepackage{graphicx}
\usepackage{dcolumn}
\usepackage{bm}
\usepackage{float}
\usepackage{subcaption}
\usepackage{dsfont}
\usepackage{multirow}
\usepackage{tikz}
\usepackage{listofitems}
\usetikzlibrary{arrows.meta}
\usepackage[outline]{contour}
\contourlength{1.4pt}

\usepackage{xcolor}
\colorlet{custom_red}{red!85!black}
\colorlet{custom_blue}{blue!95!black}
\colorlet{custom_green}{green!85!black}
\colorlet{custom_orange}{orange!70!red!80!black}
\colorlet{custom_dark_blue}{blue!10!black}
\colorlet{custom_light_blue}{blue!50!white}

\tikzset{
    >=latex,
    node/.style={thick,circle,draw=custom_blue,minimum size=22,inner sep=0.5,outer sep=0.6},
    rectangle_node/.style={thick,rectangle,draw=custom_blue,minimum size=22,inner sep=0.5,outer sep=0.6},
    node in/.style={node,green!20!black,draw=custom_green!30!black,fill=custom_green!25},
    node hidden/.style={node,blue!20!black,draw=custom_blue!30!black,fill=custom_blue!20},
    node planewave/.style={node,orange!20!black,draw=custom_orange!30!black,fill=custom_orange!20},
    node out/.style={node,red!20!black,draw=custom_red!30!black,fill=custom_red!20},
    node loss/.style={rectangle_node,blue!40!black,draw=custom_dark_blue!30!black,fill=custom_dark_blue!20},
    connect/.style={thick,custom_dark_blue},
    connect arrow/.style={-{Latex[length=4,width=3.5]},thick,custom_dark_blue,shorten <=0.5,shorten >=1},
}

\newcommand{\figref}[2][\figurename~]{#1\ref{#2}}

\newcommand{\secref}[2][Section~]{#1\ref{#2}}

\begin{document}

\preprint{APS/123-QED}

\title{Plane-Wave Decomposition and Randomised Training; a Novel Path to Generalised PINNs for SHM}

\author{Rory Clements}
 \affiliation{Faculty of Environment, Science and Economy, \\University of Exeter}
 
\author{James Ellis}
 \email{james.ellis@oxinst.com}
\affiliation{
 Oxford Instruments Plasma Technology, Bristol, BS35 4GG, UK   
}

\author{Geoff Hassall}
 \email{geoff.hassall@oxinst.com}
\affiliation{
 Oxford Instruments Plasma Technology, Bristol, BS35 4GG, UK 
}
\author{Simon Horsley}
\author{Gavin Tabor}
\affiliation{
 Faculty of Environment, Science and Economy\\
 University of Exeter
}

\date{\today}

\begin{abstract}
In this paper, we introduce a formulation of Physics-Informed Neural Networks (PINNs), based on learning the form of the Fourier decomposition, and a training methodology based on a spread of randomly chosen boundary conditions. By training in this way we produce a PINN that generalises; after training it can be used to correctly predict the solution for an arbitrary set of boundary conditions and interpolate this solution between the samples that spanned the training domain. We demonstrate for a toy system of two coupled oscillators that this gives the PINN formulation genuine predictive capability owing to an effective reduction of the training to evaluation times ratio due to this decoupling of the solution from specific boundary conditions.
\end{abstract}

\maketitle

\section{\label{sec:introduction}Introduction\protect\\}
Physics-Informed Neural Networks (PINNs) \cite{raissi2019physics} can be used as an AI-based approach to solving mathematical and physical problems, where a comprehensive, domain-spanning training dataset is not utilised \cite{mao2020physics}. The broader category of Artificial Neural Networks (ANN) are tasked with mapping an independent input training dataset, $X$, with a dependent output training dataset, $Y(X)$, by minimising a loss function that is often based on a least-squares fit to the training data. This means a successfully trained ANN represents an approximation of the map between the $X$ and $Y$ data in the training and testing datasets. Unlike ANNs, it is possible to train a PINN without any external training data at all. This may be done by operating on the solution generated by the PINN such that the governing equations of the system being investigated are in terms of the solution that has been approximated by the network. The closer the PINN is to the actual solution, the better the governing equations are satisfied. In this use case, only one part of the ``training'' dataset is required; a distribution of independent variables, such as time or position that have been sampled from their parent spaces. The loss function can then be based, in whole or in part, on the underlying expression being solved, thus ensuring the primacy of the physical laws governing the system, as well as offering a way to eliminate the need for domain-spanning training data, which is often costly to obtain. Beyond just being an alternative to existing solvers like SciPy's \texttt{odeint()} \cite{2020SciPy-NMeth}, this method might be adapted to solve equations such as Navier-Stokes without expensive training data generated by existing computational fluid dynamics methods. 

In general, in order for a PINN to solve a differential equation it must minimise a combination of the errors that arise from substituting its solution into said differential equations and the associated boundary conditions, ultimately learning a single specific solution to the problem. Because of this, the time taken to train the network cannot be axiomatically separated from the time taken to evaluate it. Thus, PINNs would not appear to be competitive with traditional numerical methods for achieving the same solution (by a very wide margin, see Table \ref{training_and_results} for the times taken using a PINN and \texttt{odeint()} to solve the same system). Our aim in this paper is to demonstrate that a PINN may be trained in a more generalised fashion across a range of input boundary conditions, creating a ``solution engine'' of sorts, that can then be used to solve a given system for a range of arbitrary boundary conditions much faster than \texttt{odeint()} can, once in a trained state. See \cite{nakamura2021physics} and \cite{schafer2022generalization} for related work towards this goal.

ANNs are well-known for their ability to upscale data \cite{dong2015image}. By a similar token, the PINNs featured in this work demonstrate the ability to interpolate between boundary condition samples used for training, meaning that the trained PINN can solve for a continuum of boundary conditions inside a certain range. After enough evaluations of the trained network for different boundary conditions, the effective training time per solution will tend towards the time taken to evaluate the trained network -- thereby reducing the training time per solution. Other work using data-driven techniques of boundary to domain mapping has been done, such as the lifting-product FNO featured in \cite{kashi2024learning}. It should be noted that in our paper we do not use externally sourced training and testing datasets for any of the neural networks. The networks featured in this work only require an input domain consisting of a set of times and an accompanying set of boundary conditions. The output from the network is automatically evaluated and compared to the differential equations, ultimately yielding a loss value. This might be thought of as on-the-fly dataset generation, without the need of an external solver to source it.

Using PINNs in this way reflects the conventional view in mathematical physics where equations such as Laplace, Poisson etc, may exhibit a general solution, which is only made unique when it must also satisfy a complete set of boundary conditions. To match this paradigm, the PINN needs to learn the general solution of the expression it is trained on, and then particularise the solution for a given set of boundary conditions only upon evaluation.

Within the subject of deep learning, network architecture (typically referring to the number and configuration of layers) is arguably one of the most important factors for how successful a network will be at a given task. Often, suitable network architectures are found by trial and improvement. In the case of a PINN, prior mathematical insight may be used to influence this architecture, such that improvements to the rate of convergence on the solution can be made. In particular we will show that for a toy system of two coupled oscillators, a plane-wave activation function (the non-linear function relating inputs and output of a neuron) is beneficial in the training and solution process.

It should be noted that there are other methods of employing Fourier and Fourier-like transformations in neural network models. A notable example is that of the Fourier Neural Operator (FNO) \cite{li2020fourier} and \cite{kovachki2023neural}. Another body of work \cite{pribotkin2023using} demonstrates that the use of trigonometric activation functions (other than $\tanh$) can be beneficial when solving a system of two coupled Duffing oscillators. Other work on the use of atypical activation functions includes \cite{abbasi2024physical}, where they introduce ``Physical activation functions'', and how they can help reduce the size of PINNs. Finally, and of relevance to this paper is \cite{cui2022efficient} where plane-wave activation functions are used to solve the Helmholtz equation.

The rest of this paper is structured as follows. Before discussing the methodology and results, a succinct coverage of ANNs (section \ref{sec:ANNs}) is made, whilst the key concepts of ``pure physics'' PINNs (those that rely solely on differential equations for training, not supplemented by externally sourced training data) are discussed in section \ref{sec:PINNs}. The test system used in this work is that of two coupled oscillators, and is introduced in section \ref{sec:lossFunction}; this was chosen because it can be used as a simplified model for many physical systems. Different types of activation functions are introduced in \secref{sec:conventional_activation_functions} and \secref{sec:plane_wave_PINN_theory}, with a distinction made between the popular choices such as $\tanh$ (we refer to these as conventional activation functions as they are not necessarily representative of a meaningful decomposition of the expected solutions to the system being solved) and more physically relevant activation functions drawn from a knowledge and understanding of the mathematics of the expected solution, such as plane waves. Section \ref{sec:PINNtraining} presents the method by which the PINNs are trained to generalise for a range of input boundary conditions. The results are presented in section \ref{results} for three types of PINN; with conventional activation function, plane-wave PINNs and generalised plane-wave PINNs. In each case the PINN solution is compared with numerical solutions generated using the {\tt odeint()} function from the SciPy library, as well as an analysis and comparison between the PINN solutions themselves. The results of the paper are summarized in the Conclusions section (section \ref{conclusions}). Finally, Appendix \ref{Append_A} covers further results for conventional PINNs, where the number of trainable parameters is a closer match to the number present in the plane-wave PINNs featured in the body of this paper.

\subsection{Artificial Neural Networks} \label{sec:ANNs}
In general, the task of a neural network is to determine the mapping between a set of dependent variables (Y data) to a set of independent variables (X data). Broadly, this ability of neural networks is because they are universal function approximators \cite{hornik1989multilayer}. Training ANNs typically requires large quantities of externally sourced training data, which is often contaminated with noise or exemplifies characteristics other than those that the network should learn. A high ratio of noise to meaningful information in the training dataset can sometimes require a lower learning rate, in order for the network to learn a reasonable generalisation. Despite these difficulties, ANNs are being applied to nearly every scientific discipline today, with varying degrees of success. In this paper, we will demonstrate that they are also of use in the theoretical domain, where they might help shed some light on solutions to previously intractable problems. What follows is a lightweight introduction to artificial neural networks. For more detailed coverage of the theory behind deep learning in general, see \cite{roberts2022principles}.

\begin{figure}
\begin{tikzpicture}[x=1.8cm,y=1.4cm]
  \message{^^ANN}
  \readlist\Nnod{1,5,5,1,1}
  \def\NH{4}
  \def\nstyle{int(\lay<\Nnodlen?(\lay<\NH?min(2,\lay):3):4)}
  \tikzset{
    node 1/.style={node in},
    node 2/.style={node hidden},
    node 3/.style={node out},
    node 4/.style={node loss},
  }
  \message{^^J  Layer}
  \foreachitem \N \in \Nnod{ 
    \def\lay{\Ncnt} 
    \pgfmathsetmacro\prev{int(\Ncnt-1)} 
    \message{\lay,}
    \foreach \i [evaluate={\y=\N/2-\i; \x=\lay; \n=\nstyle;}] in {1,...,\N}{ 
      
      \node[node \n] (N\lay-\i) at (\x,\y) {};
      
      \ifnum\lay>1 
        \foreach \j in {1,...,\Nnod[\prev]}{ 
          \draw[connect,white,line width=1.2] (N\prev-\j) -- (N\lay-\i);
          \draw[connect] (N\prev-\j) -- (N\lay-\i);
        }
      \fi 
    }
  }
  \node[above=15,align=center,custom_green!60!black] at (N1-1.90) {Input\\[-0.2em]X data};
  \node[above=2,align=center,custom_blue!60!black] at (N2-1.90) {Hidden layers};
  \node[above=15,align=center,custom_red!60!black] at (N4-1.90) {Network\\approximation\\[-0.2em]for\\$Y(X)$};
  \node[above=8,align=center,custom_dark_blue!60!black] at (N\Nnodlen-1.90) {Loss\\[-0.2em]function\\$L(X,Y)$};
\end{tikzpicture}
\caption{A depiction of an archetypal fully connected artificial neural network (ANN), where the coloured circles represent individual neurons. Layers are represented by vertical stacks of neurons. The training data is supplied to the network at the left- and right-most nodes from the $X$ and $Y$ datasets respectively, where $X$ is the set of independent data and $Y(X)$ is the set of dependent data. It is the task of the ANN to ``learn'' the mapping $Y(X)$ between the training datasets.}
\label{fig:fig_ANNs_100}
\end{figure}
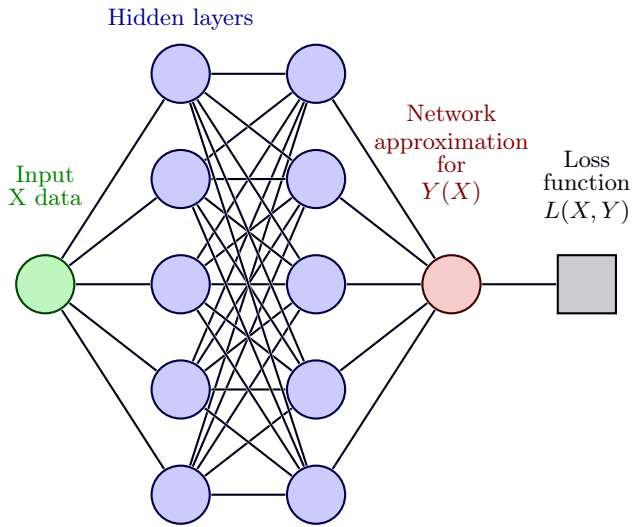

At the heart of any neural network is the artificial neuron, which is a choice of non-linear function applied to a linear sum of inputs. Multiple neurons are arranged in parallel to form a layer, and many layers are stacked together to form a neural network. If every neuron is connected to every other in the next layer then the network is referred to as being fully connected (which broadly falls into the category of being a multilayer perceptron \cite{rosenblatt1958perceptron}), depicted by \figref{fig:fig_ANNs_100} \cite{NeutelingsNNfigs}.
The action of a neuron in an ANN (such as that depicted by \figref{fig:fig_ANNs_100}) is best described explicitly for the $n^\text{th}$ neuron in the $l^\text{th}$ layer. The output activation $a_n^l$ for this neuron is described by the expression
\begin{equation} \label{eq:layer_function}
    a_n^l = \sigma^l \left( \sum_i \text{w}_{ni}^l a_{i}^{l-1} + b_n^l \right) = \sigma^l (z_n^l)
\end{equation}
where $\sigma^l$ is a differentiable, nonlinear activation function that is applied to all of the neurons in the layer $l$. $\text{w}_{n,i}^{l}$ is the $i^\text{th}$ ``weight'' associated with the $n^\text{th}$ neuron in layer $l$, $a_{i}^{l-1}$ is the output activation from the $i^\text{th}$ neuron in the preceding layer $l-1$. $b_n^l$ is the additive constant or ``bias'' for the neuron in the $l^\text{th}$ layer. Finally, $z^l$ is the sum of linear functions in the layer $l$ before application of the activation function $\sigma^l$. $\text{w}_{ni}^l$ and $b_n^l$ are known as the trainable parameters of the network, since it is these that the backpropagation process (see below) is able to adjust in order to improve the convergence that the network has on the mapping. The summation over all $i$ is specific to fully connected networks, where every neuron in the layer $l$ is connected to every neuron in the layer $l-1$ and $l+1$. This is one of the simplest and most ubiquitous ways of constructing inter-layer relationships within a neural network, but there are many other popular methods too, often with increasing levels of complexity. However, this paper will be using only fully connected architectures.

Training an ANN or PINN typically requires a loss function, represented by the square node in \figref{fig:fig_ANNs_100}. The purpose of this function is to quantify how effective the neural network is at approximating the mapping $Y(X): X \rightarrow Y$. A simple but effective function used for ANNs is least-squares, $L = (1/N) \sum_i^N [Y_{i;\text{true}} - Y_{i;\text{network}}(X)]^2$ where $L$ is the loss, $Y_\text{true}$ is the known $Y$ data from the training set, $Y_\text{network}$ is the network's prediction for the $Y$ data, and $N$ is the total number of samples present in the $X$ and $Y$ datasets. The direction and amounts by which the trainable parameters in the network should be adjusted can be determined from the scalar value the loss function returns, in order to improve the network's convergence on the mapping. This is performed using the backpropagation process. The full details of backpropagation and its operation is outside of the scope of this paper, but it is informative to provide a cursory analysis of the process. See \cite{Aggarwal:2023} for more detailed coverage of backpropagation and \cite{rumelhart1986learning} for one of the original papers on the subject.

By determining the loss function as a function of the trainable parameters in the network (in addition to the real $Y$ training data, $Y_\text{true}$ and the network's approximation of the $Y$ data $Y_\text{network}$), $L(Y_\text{true}, Y_\text{network}, \text w_{ni}^l, b_n^l)$, and assuming the first derivative exists for the loss and activation functions used throughout the network, adjustments to the weights $\text{w}_{ni}^l$ and biases $b_n^l$ by amounts such that $L$ may be reduced upon another evaluation of the network can be made. Using the chain rule, the first derivative of the loss with respect to every weight and bias in the entire neural network may be determined, and adjusted accordingly. See \cite{Aggarwal:2023} for further coverage on this topic and the precise method by which the chain rule is applied.

As mentioned above, the loss value required for the backpropagation process is obtained by performing an evaluation or a forward propagation through the network. The backpropagation process described so far allows for more than one way of updating the trainable parameters. For example, one could choose to update a single weight or bias using the loss value obtained from the forward propagation, then evaluate the network again such that another might be updated. This would result in a fairly direct path towards a local or global minima on the loss landscape. The loss landscape is the high-dimensional surface defined by the loss function, existing in a dimension equivalent to the number of trainable parameters in the network. Explicitly, this dependence is given by $L = (1/N) \sum_i^N [Y_\text{true} - Y_\text{network}(X, \text{w}_{ni}^l, b_n^l)]^2$ for a least-squares loss, where $Y_\text{network}(X, \text{w}_{ni}^l, b_n^l)$, is the mapping as described by the network. In addition to relating the $Y(X)$ output of the network to the $X$ training data it also relates it to all of the trainable parameters in the network. The loss landscape might be imagined as a potential that the system point (existing in the space as defined by the trainable parameters in the network) is experiencing. In general, it is more likely that the system point will get stuck in a local rather than global minimum, but in practice a deep enough local minimum usually represents a configuration of the network that yields a suitable approximation of the desired mapping. The problem with this approach is that a typical neural network may contain thousands (or many, many more) of trainable parameters, and so training the network by rigorously updating a single parameter and then re-evaluating is extremely time-consuming. Instead, a more stochastic process is used. A single forward propagation through the network is used to calculate the loss, and then all the gradients associated with the trainable parameters are computed and updated by a small amount simultaneously. Naturally, this results in the system point undergoing more of a random walk rather than the uniform gradient descent towards the local or global minima that might be had by rigorously changing one parameter at a time. By choosing a low enough learning rate (a multiplicative constant applied to the gradient of the loss with respect to the weight or bias in question), the system point will fairly reliably find a minima. See \cite{choromanska2015loss} and \cite{li2018visualizing} for a much more in-depth discussion on optimisation and visualisation of loss functions and landscapes.

\subsection{Physics-Informed Neural Networks}\label{sec:PINNs}
One of the main disadvantages of neural networks is that they typically require vast quantities of externally sourced training data. This is partly due to the low learning rate used to scale the gradients obtained during each backpropagation step. This low learning rate parameter is to allow the system point in the configuration space defined by the trainable parameters to be more strongly influenced by the ``potential'' from the loss landscape than the random walk induced by the stochasticity of the backpropagation process. That is, if the magnitude of the random walk is larger than the magnitude of the ``force'' experienced by the system point due to the local gradients of the loss landscape, then the network is unlikely to optimise. As a result, a large quantity of training data is required for the network to converge on the desired mapping. It should be noted that excessive exposure to the same small training data set can lead to its own problems, such as a failure to form a good generalisation. In the field of physics, such training data often comes from simulation or experiment. Unfortunately, data obtained in this way in the quantities required to train a neural network is often prohibitively expensive or time-consuming to generate. This is where a Physics-Informed Neural Network or ``PINN'' can be of benefit.

A simple PINN may be seen as a traditional ANN but with a modified loss function, either additional to or replacing the least-squares loss used for determining the map between externally sourced $X$ and $Y(X)$ training data. PINNs of this type are able to be trained without externally sourced training, essentially making them into a solver for differential equations. In the context of this paper, we use the phraseology ``pure physics'' to distinguish these PINNs from those that apply differential equations to the loss function in order to accelerate the abstraction of a map from externally sourced training data. These ``pure physics'' PINNs are very similar in design and function to the PINN used for the forward problem of the Euler equations in \cite{mao2020physics}.

It is possible to impose any number of additional governing laws of physics or mathematical properties that a system is expected to exhibit onto a PINN in this way, rather than assuming that an ANN will abstract them from externally sourced training data alone. Such constraints may be equations of motion, boundary conditions, conserved or invariant quantities, continuity etc. These constraints may be reformulated into new terms in the loss function, and will modify the loss landscape accordingly, such that the resulting geometry better attracts the system point to a local or global minima. This can enable a PINN to train faster than an ANN and exhibit better overall convergence on the desired mapping using less or no externally sourced training data. As mentioned above, we shall not be considering PINNs that utilise externally sourced training data in the loss function, but instead focus on solving the differential equations that model the dynamics of a toy system of two coupled oscillators, through the use of the aforementioned ``pure physics'' PINNs. It should also be noted that unless the noise associated with this training data is well modelled in the governing equations, the resulting loss landscape formed by the training data and governing equations can be contradictory, resulting in an unexpected mapping learned by the network. This is not a problem that ``pure physics'' PINNs suffer from.

The aforementioned modifications to the loss function required to implement a PINN are best shown by example. Consider the differential equation for damped harmonic motion, $(m\partial_t^2 + b\partial_t + k)x=0$ where $m$, $b$, $k$ and $x$ represent the mass, damping and spring constants and position respectively. In order to solve this expression, the time domain is passed to the ``Input $X$ data'' layer, and the PINN will then generate an approximation for the position domain in the ``Network approximation for $Y(X)$'' layer upon a feedforward evaluation of the network. We may adjust the ANN depicted by \figref{fig:fig_ANNs_100} such that it might solve this expression, purely by modifying the loss function. For damped harmonic motion, the loss function is given by $$L_\text{EOM} = \frac{1}{N} \sum_{j=0}^{N}\left[\left(m\frac{\partial^2}{\partial t^2} + b \frac{\partial}{\partial t} + k\right)x_P(t_j)\right]^2$$
where $x_P(t_j)$ is the PINN solution for the $x$ coordinate at time step $t_j$. The summation over $j$ is to ensure that the equation of motion is adhered to for all the discrete time steps between the initial and final times. A PINN capable of minimising such a loss to a suitably low value indicates that a solution for the equation of motion $x_P(t)$ has been successfully modeled. Much more in-depth coverage of PINNs and many other areas in machine learning can be found in \cite{Brunton:2022}.

The remaining components of ANNs and PINNs that are of key importance to the rest of this paper are the activation functions. As discussed above, these functions must be differentiable and are typically nonlinear. Unlike the majority of use cases for ANNs, PINNs enjoy the luxury of solving problems that we have a chance of understanding (if not solving) analytically. This presents the possibility of being able to make inroads on selecting activation functions that are more meaningful (if perhaps not perfect) decompositions of the expected solutions. It is possible to impose such decompositions by noting that for a fully connected network architecture, the input to a neuron in the $l^\text{th}$ layer is the summation of outputs from the neurons in the layer $l-1$. The utilisation of such ``informed choices'' of activation functions is at the heart of creating PINNs that can be trained to solve problems with previously unseen boundary conditions, and is the subject of this paper.

\section{Method\protect\\} \label{sec:method}
As mentioned previously, attention is focused specifically on solving a toy system of two coupled oscillators using the PINNs featured in this paper. The motivation for this choice is due to its linearity, step up in complexity from the single oscillator and applicability to quantum and solid state physics. The method used for our investigations comes down to a comparison between two architecturally different types of PINN. The first, and most recognisable, is a fully connected PINN using conventional activation functions. The second is the plane-wave PINN. At its most fundamental, the plane-wave PINNs featured in this work modify the fully connected PINN architecture to include at least one layer that uses plane-wave activation functions. The point of this layer is to produce insightful decompositions of the expected solutions when a summation is made over the whole layer by the following one. It should be noted that the use of plane-wave/unconventional trigonometric activation functions is not novel to this work, see \cite{pribotkin2023using}, \cite{abbasi2024physical}, \cite{cui2022efficient} and \cite{raissi2019deep}. However, before discussing the network architectures further, the loss function should first be derived.

\subsection{Constructing the loss function}\label{sec:lossFunction}
\begin{figure}
	\centering
	\includegraphics[width=75mm]{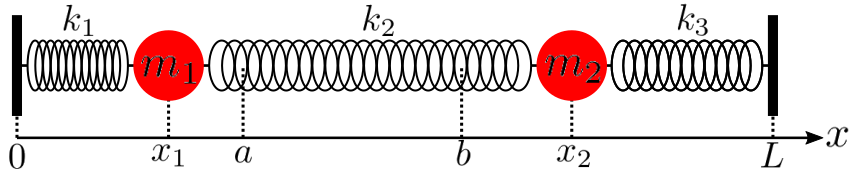}
	\caption{A system of two coupled oscillators, with rigid boundaries at $x=\{0,L\}$. $a$ and $b$ represent the equilibrium positions of the masses. $x_1$ and $x_2$ represent the displacements of the oscillators $m_1$ and $m_2$ from their equilibrium positions respectively. $k_1$, $k_2$ and $k_3$ represent the spring constants associated with each spring in the system.}
	\label{fig:fig_method_100}
\end{figure}

Systems of coupled oscillators are used widely throughout physics for modeling multipartite systems where the Hamiltonian depends upon significant energy contributions arising from inter-particle relationships. The very simplest of these systems is two coupled oscillators, depicted by \figref{fig:fig_method_100}.
Typically, loss functions need to have a lower bound, i.e, they should not exhibit collapse states where the loss can become infinitely negative. In a PINN, this is typically achieved by taking the mean square of Equations \eqref{eq:COEOM1} and \eqref{eq:COEOM2}, such that the correct mapping between time and displacements $x_1$ and $x_2$ produces the smallest (zero) loss value. The equations of motion for a system of two coupled oscillators is given by
\begin{align}
			m_1\ddot{x}_1 + m_1\omega_1^2 x_1 - k_2 x_2 &= 0 \label{eq:COEOM1}\\
            m_2\ddot{x}_2 + m_2\omega_2^2 x_2 - k_2 x_1 &= 0 \label{eq:COEOM2}
\end{align}
where $x_{1,2}$ are the displacements from $a$ and $b$ respectively, and the frequencies $\omega_{1,2}$ are given by
\begin{align*}
    \omega_1^2 &= \frac{k_1+k_2}{m_1}\\
    \omega_2^2 &= \frac{k_2+k_3}{m_2}
\end{align*}
These two expressions can be expressed as terms in the loss function for the PINN by taking the mean square of the left-hand sides
\begin{equation}
			L_1 = \frac{1}{N}\sum_{j=0}^N\left[m_1\ddot{x}_{1,P}(t_j) + m_1\omega_1^2 x_{1,P}(t_j) - k_2 x_{2,P}(t_j)\right]^2 \\ \label{eq:COL1}
\end{equation}
\begin{equation}
            L_2 = \frac{1}{N}\sum_{j=0}^N\left[m_2\ddot{x}_{2,P}(t_j) + m_2\omega_2^2 x_{2,P}(t_j) - k_2 x_{1,P}(t_j)\right]^2 \label{eq:COL2}
\end{equation}
where $N$ represents the number of time steps, and $x_{\{1,2\},P}$ are the network outputs for the positional coordinates $x_{\{1,2\}}$. The mean over all the time steps is to ensure that the loss terms incorporate contributions from the solutions for $x_{1,P}(t)$ and $x_{2,P}(t)$ over the entire training time domain. The overall loss function for the dynamics is the sum of these two loss terms, $L_\text{dynamics} = L_1 + L_2$ and represents a quantitative measure of how well the PINN is mapping the time domain to the position domains $x_1$ and $x_2$. These loss terms by themselves are not sufficient to restrict the PINN to a unique solution. In order to do this, one must impose the boundary conditions for the coordinates $x_1$ and $x_2$ as well. A set of boundary condition loss terms are given below
\begin{align*}
			L_3 &= \left[x_{1,P}(t_0) - x_{1,0} \right]^2 \\
            L_4 &= \left[\dot{x}_{1,P}(t_0) - v_{1,0} \right]^2 \\
            L_5 &= \left[x_{2,P}(t_0) - x_{2,0} \right]^2 \\
            L_6 &= \left[\dot{x}_{2,P}(t_0) - v_{2,0} \right]^2
\end{align*}
where $x_{\{1,2\},P}(t_0)$ and $\dot{x}_{\{1,2\},P}(t_0)$ are the PINN solution's for position and velocity of the oscillators at the initial time step, $j=0$. $x_{\{1,2\},0}$ and $v_{\{1,2\},0}$ are the boundary conditions for position and velocity of the oscillators at the initial time step $t_0$.

Typically, these loss terms are scaled by an additional multiplicative hyperparameter, to enable some user control over the loss landscape during the training process. For the system of coupled oscillators, the final loss function has the form
\begin{equation}
    \label{eq:loss}
    L = \Lambda_1(L_1 + L_2) + \Lambda_3(L_3 + L_5) + \Lambda_3(L_4 + L_6)
\end{equation}
where $\Lambda_n$ are the hyperparameters scaling each loss term. The grouping of position and velocity loss terms is done such that each of the loss weighting hyperparameters scale the dynamics, positional and velocity boundary conditions for each oscillator respectively.

\subsection{Conventional PINNs}
\label{sec:conventional_activation_functions}
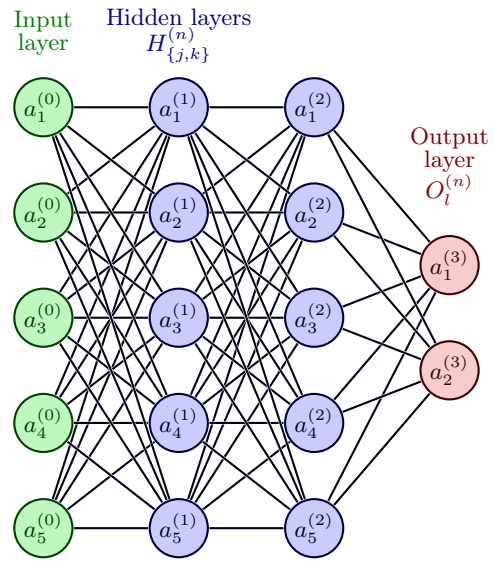
\begin{figure}
\begin{tikzpicture}[x=1.8cm,y=1.4cm]
  \message{^^Plane-wave PINN}
  \readlist\Nnod{5,5,5,2}
  \def\NH{4}
  \def\nstyle{int(\lay<\Nnodlen?(\lay<\NH?min(2,\lay):3):4)}
  \tikzset{
    node 1/.style={node in},
    node 2/.style={node hidden},
    node 4/.style={node out},
  }
  \message{^^J  Layer}
  \foreachitem \N \in \Nnod{ 
    \def\lay{\Ncnt} 
    \pgfmathsetmacro\prev{int(\Ncnt-1)} 
    \message{\lay,}
    \foreach \i [evaluate={\y=\N/2-\i; \x=\lay; \n=\nstyle;}] in {1,...,\N}{ 

      \node[node \n] (N\lay-\i) at (\x,\y) {$a_\i^{(\prev)}$};

      \ifnum\lay>1 
        \foreach \j in {1,...,\Nnod[\prev]}{ 
          \draw[connect,white,line width=1.2] (N\prev-\j) -- (N\lay-\i);
          \draw[connect] (N\prev-\j) -- (N\lay-\i);
        }
      \fi 
    }
  }
  \node[above=5,align=center,custom_green!60!black] at (N1-1.90) {Input\\[-0.2em]layer};
  \node[above=2,align=center,custom_blue!60!black] at (N2-1.90) {Hidden layers\\[-0.2em]$H_{\{j,k\}}^{(n)}$};
  \node[above=8,align=center,custom_red!60!black] at (N\Nnodlen-1.90) {Output\\[-0.2em]layer\\$O_l^{(n)}$};
\end{tikzpicture}
\caption{Network architecture for a conventional PINN. Apart from the input and output layers, the architecture depicted here is representative only. }
\label{fig:conventional_PINNs_100}
\end{figure}

As discussed in section \ref{sec:ANNs}, ANNs typically utilise non-linear activation functions, such as sigmoid, $\tanh$, ReLU \cite{householder1941theory} and many others. These functions are often chosen by trial and improvement, depending on the network architecture and the mapping to be modeled by the network. Throughout this work we refer to these as conventional activation functions and networks, since they are very popular, mainly due to how well they work for a wide variety of networks. This distinction also helps to differentiate these activation functions and networks from those chosen to, in some way, reflect the expected solutions. The general form of the network architecture used for the conventional PINNs in this work is depicted by \figref{fig:conventional_PINNs_100} \cite{NeutelingsNNfigs}, and uses $\tanh$ activation functions only. This figure demonstrates that the output layer will represent solutions as a summation of the outputs from the previous, hidden layer. Since the conventional networks featured in this paper use $\tanh$ activation functions, the position-domain solutions are represented as a summation of many $\tanh$ functions for each time step in the time domain.
It is instructive to decompose the network depicted by \figref{fig:conventional_PINNs_100} into discrete mathematical steps, shown by \eqref{eq:conventional_walkthrough_1}, \eqref{eq:conventional_walkthrough_2} and \eqref{eq:conventional_walkthrough_3}. It should be noted that each step representing the output from each layer has been separated, such that the indices of repeated summations do not become intractable.
\begin{align}
    H_j^{(1)} &= \tanh \left[ \sum_i^{a^{(0)}} \left(\text w_{ji}^{(1)} a_i^{(0)} \right) + b_j^{(1)} \right] \label{eq:conventional_walkthrough_1} \\
    H_k^{(2)} &= \tanh \left[ \sum_j^{a^{(1)}} \left(\text w_{kj}^{(2)} H_j^{(1)} \right) + b_k^{(2)} \right] \label{eq:conventional_walkthrough_2} \\
    O_l^{(3)} &= \tanh \left[ \sum_k^{a^{(2)}} \left(\text w_{lk}^{(3)} H_k^{(2)} \right) + b_l^{(3)} \right] \label{eq:conventional_walkthrough_3}
\end{align}
By representing the layers separately in this way the indices still require further clarification. For a given layer $l$, each neuron $j$ will have its own set of weights $w_{ji}^{(l)}$ that scale the outputs from the previous layer $a_i^{(l-1)}$. This is why the weights are given as a matrix.

\subsection{Plane-Wave PINNs}\label{sec:plane_wave_PINN_theory}
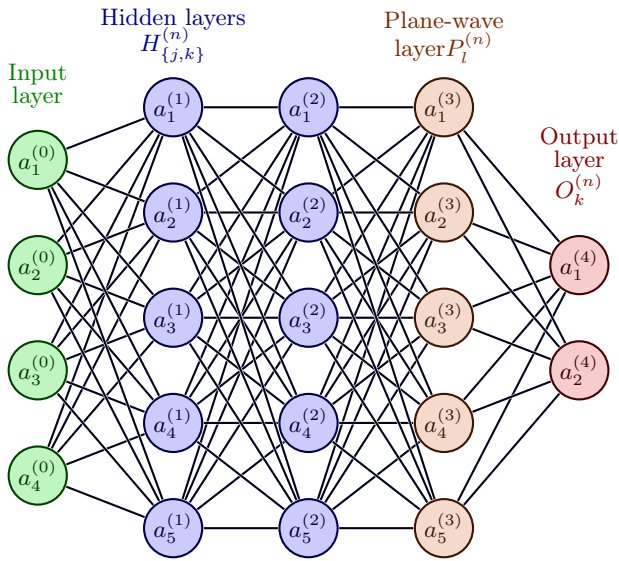
\begin{figure}
\begin{tikzpicture}[x=1.8cm,y=1.4cm]
  \message{^^Plane-wave PINN}
  \readlist\Nnod{4,5,5,5,2}
  \def\NH{4}
  \def\nstyle{int(\lay<\Nnodlen?(\lay<\NH?min(2,\lay):3):4)}
  \tikzset{
    node 1/.style={node in},
    node 2/.style={node hidden},
    node 3/.style={node planewave},
    node 4/.style={node out},
  }
  \message{^^J  Layer}
  \foreachitem \N \in \Nnod{ 
    \def\lay{\Ncnt} 
    \pgfmathsetmacro\prev{int(\Ncnt-1)} 
    \message{\lay,}
    \foreach \i [evaluate={\y=\N/2-\i; \x=\lay; \n=\nstyle;}] in {1,...,\N}{ 

      \node[node \n] (N\lay-\i) at (\x,\y) {$a_\i^{(\prev)}$};

      \ifnum\lay>1 
        \foreach \j in {1,...,\Nnod[\prev]}{ 
          \draw[connect,white,line width=1.2] (N\prev-\j) -- (N\lay-\i);
          \draw[connect] (N\prev-\j) -- (N\lay-\i);
        }
      \fi 
    }
  }
  \node[above=5,align=center,custom_green!60!black] at (N1-1.90) {Input\\[-0.2em]layer};
  \node[above=2,align=center,custom_blue!60!black] at (N2-1.90) {Hidden layers\\[-0.2em]$H_{\{j,k\}}^{(n)}$};
  \node[above=2,align=center,custom_orange!60!black] at (N4-1.90) {Plane-wave\\layer$P_l^{(n)}$};
  \node[above=8,align=center,custom_red!60!black] at (N\Nnodlen-1.90) {Output\\[-0.2em]layer\\$O_k^{(n)}$};
\end{tikzpicture}
\caption{Network architecture for a plane-wave PINN. Apart from the input and output layers, the architecture depicted here is representative only.}
\label{fig:fig_PWPINNs_100}
\end{figure}

As discussed above, the position-domain solutions generated by the conventional PINN (detailed in sections \ref{sec:lossFunction}, \ref{sec:conventional_activation_functions} and depicted by \figref{fig:conventional_PINNs_100}) for each time step is the summation of many $\tanh$ functions. This insight suggests summation-style decompositions of the expected solutions are likely to be good choices of activation functions. For the toy system of two coupled oscillators, the most obvious of these decompositions is a plane-wave decomposition. \figref{fig:fig_PWPINNs_100} \cite{NeutelingsNNfigs} depicts the general structure of a ``plane-wave PINN''.

Again, it is instructive to decompose the network depicted by \figref{fig:fig_PWPINNs_100} into discrete mathematical steps, shown by \eqref{eq:plane_wave_walkthrough_1}, \eqref{eq:plane_wave_walkthrough_2}, \eqref{eq:plane_wave_walkthrough_3} and \eqref{eq:plane_wave_walkthrough_4}. As was the case for the conventional PINN, it should be noted that each step representing the output from each layer has been separated such that the indices of repeated summations do not become intractable.
\begin{align}
    H_j^{(1)} &= \tanh \left[ \sum_i^{a^{(0)}} \left(\text w_{ji}^{(1)} a_i^{(0)} \right) + b_j^{(1)} \right] \label{eq:plane_wave_walkthrough_1} \\
    H_k^{(2)} &= \text{sigmoid} \left[ \sum_j^{a^{(1)}} \left(\text w_{kj}^{(2)} H_j^{(1)} \right) + b_k^{(2)} \right] \label{eq:plane_wave_walkthrough_2} \\
    P_l^{(3)} &= \exp{i \omega_l t} \left[ \sum_k^{a^{(2)}} \left(\text w_{lk}^{(3)} H_k^{(2)} \right) + b_l^{(3)} \right] \label{eq:plane_wave_walkthrough_3} \\
    O_p^{(4)} &= \left[ \sum_l^{a^{(3)}} \left(\text w_{pl}^{(4)} P_l^{(3)} \right) + b_p^{(4)} \right] \label{eq:plane_wave_walkthrough_4}
\end{align}

The plane-wave activation function applied in the plane-wave PINNs used throughout this paper is given by
\begin{equation}
    \label{eq:PWActivation}
    \sigma = z \exp{\mathrm i \omega t}
\end{equation}
where $z$ is the output from the linear equation, defined by \eqref{eq:layer_function}, $\omega$ is the angular frequency associated with a given plane-wave, and $t$ is the time domain over which the PINN is being trained. Note that the indices have been omitted in \eqref{eq:PWActivation} for clarity. It should also be noted that the activation function itself relies on an additional pair of independent variables, $\omega$ and $t$, of which $\omega$ represents another degree of freedom, and this may be adjusted for further tuning of the network. Naturally, $\omega$ and $t$ are arrays of variables, passed to the activation function such that each of the neurons in the plane-wave layer $P_l^{(3)}$ contributes a unique plane-wave to the output layer $O_p^{(4)}$. This unique plane-wave corresponds to the angular frequency in the array of $\omega$ that was used for that particular neuron. Since each plane wave is different due to using a unique angular frequency $\omega_l$, it might be said that the plane-wave layer is actually many independent activation functions. It should be noted that the plane-wave PINNs used throughout this paper only use one plane-wave layer, situated penultimately. The last layer's task is completing the evaluation of the plane-wave solution through summation of the plane-wave components, with no further activation function being applied at this point.

\subsection{PINN Training}\label{sec:PINNtraining}
In this work, the training of the networks is somewhat atypical. The training and evaluation of both the conventional and plane-wave PINNs is first performed using a fixed set of boundary conditions. Following this, decoupled plane-wave PINNs are explored, where the PINN mapping is not dependent on a unique set of boundary conditions. This is achieved by training the network on a different set of boundary conditions for each training step, where the boundary conditions are sampled at random from normal distributions. Then the PINN's performance is quantified by evaluating it on a fixed set of boundary conditions. These boundary conditions represent the mean values for each of the normal distributions that were used to sample each boundary condition from during training. In order to get a better idea of how well the PINN abstracts away from the mean values, a range of boundary conditions was used to evaluate the plane-wave PINNs as well, discussed further in \secref{sec:verification} and \secref{sec:GSPWPINNs}.

\subsection{Decoupling the Solution from the Boundary Conditions} \label{sec:decoupled_bc_theory}
As mentioned in \secref{sec:plane_wave_PINN_theory}, the plane-wave activation function given by \eqref{eq:PWActivation} is dependent on angular frequencies $\omega$ and times $t$. The introduction of time at this point in the network means that the independent variables passed to the input layer of the PINN can be used for another purpose. The simplest approach is to train the network with random noise; thereby decoupling the input layer from the output entirely. So far as the network is concerned, the random noise input only serves to stimulate different outputs, which is important during training such that the same activations and resulting loss values are not repeatedly used for backpropagation. To some extent, the input of the network is now a ``free'' set of parameters, and can be put to use. Rather than using noise, random sets of boundary conditions can be used instead. Without any further modifications to the PINN, it will learn to ignore these as well. However, by modifying the loss function so that the same random boundary conditions are used for each training instance, a new relationship between the boundary conditions and general solution is formed in the network. The result is a trained PINN that can solve a given system for previously unseen boundary conditions (within the same normal distributions from which the training boundary conditions were sampled). Now the map the network has learned is decoupled from a single, unique set of boundary conditions. This may represent a huge reduction in the computational cost associated with solving systems using such a PINN, since a single training session yields a general solution that is made unique only when applying the boundary conditions. It should be noted that no special effort has been made to ensure that every training step uses a new set of boundary conditions. It has been assumed that the random sampling from the normal distributions used for the boundary conditions serves this purpose adequately. Since the plane-wave PINNs in this paper are a function of the boundary conditions, it is also necessary to make the conventional PINNs (detailed in \secref{sec:conventional_activation_functions}) a function of the boundary conditions. To do this, the input layer of the conventional PINN depicted by \figref{fig:conventional_PINNs_100} is passed five inputs, the four boundary conditions along with the time domain for each training step. Each of the individual boundary conditions are scalar, but the time domain is a vector, meaning that the boundary conditions must be repeated such that they each form a vector with the same number of elements as the time domain. The rest of the method by which the boundary conditions are decoupled is the same as for the plane-wave PINNs, where random samples from a normal distribution are passed to the input layer as well as the boundary condition loss terms. This method of adding a dependence on the boundary conditions to the conventional PINNs was chosen due to its relative simplicity. It is just one possible method to achieve this goal, and by no means do we mean to suggest that this is the best or only way.

\subsection{Verification and Quantification} \label{sec:verification}
Verification of results from neural networks is necessary to ensure that the results are meaningful. Unlike other methods of regression, the largely unknown way in which neural networks abstract relationships from training data makes it particularly important to check results thoroughly. In this work, we compare the PINN solutions against the ones obtained using \texttt{odeint()} from the SciPy library to numerically solve the systems, mainly due to its simplicity and ubiquity for solving differential equations, such as for the toy system of two coupled oscillators. Each PINN solution is compared both qualitatively, and through a comparison of the surfaces formed when plotting the PINN and numerical solutions over a range of boundary conditions against time. This provides both an intuition for the proficiency that a given PINN has for solving the system of coupled oscillators, along with a quantitative measure for when our judgment is no longer satisfactory. Quantification of the state of training of a neural network can also be somewhat difficult. Typically, the easiest but most abstract way to quantify how well a network has been trained is to use the most recent loss value, as discussed in \ref{sec:lossFunction} and denoted by \eqref{eq:loss}. However, in practice the rolling mean loss is a more useful quantity, due to the stochasticity of the instantaneous loss, discussed further in \secref{sec:computational_resources}.

\subsection{Non-network Parameters}
To ensure that only the network architecture was being varied independently, all other parameters have been kept constant. In addition, the evaluation boundary conditions, were also kept the same for all of the results (other than the generalised solution PINNs, for which ranges of boundary conditions were also used), as discussed in \secref{sec:decoupled_bc_theory}. This is also true for the training of PINNs that require fixed boundary conditions. For all results, the training and evaluation time domain intervals are $10^{-1}$ and $10^{-3}$ respectively. As can be seen in \secref{results}, if the positional solutions are continuous in the training time domain, they remain continuous in the evaluation time domain, indicating that the PINNs featured in this work exhibit the ability to upscale data, a property ANNs and PINNs in general exhibit, discussed in \secref{sec:introduction}. The learning rate for all solutions was set to the same value of $10^{-3}$ and the ``Adam'' \cite{kingma2014adam} optimiser was used.

\section{Results and Discussion\protect\\} \label{results}
Three main results are discussed in this paper, those of conventional, unique-solution plane-wave and generalised-solution plane-wave PINNs. In each case, the PINN and \texttt{odeint()} solutions are compared and the PINN result further analyzed. Numerical rather than anlytical solutions have been used for comparison, such that the same intervals may be used in solving for both the numerical and PINN solutions (during evaluation). As a result a meaningful comparison can be made between the PINN and numerical evaluation times for two different solvers.

\subsection{Computational Resources} \label{sec:computational_resources}
All of the training and evaluation of the follwing PINNs took place on an NVIDIA GeForce RTX 3090Ti graphical processing unit (GPU), and the \texttt{odeint()} solutions were performed on an Intel i9-12900K central processing unit (CPU) with 32GB of random access memory (RAM). In addition, all of the neural networks were constructed using the PyTorch Python module.

A full breakdown of the training conditions and associated times for training and evaluation may be found in Table \ref{training_and_results}. Each result in this paper has a reference to this table in order to provide more specific performance information about both the PINN and numerical solutions. Each PINN was trained to a point where it demonstrated a ``reasonable'' convergence on the numerical solution, where possible. This rather qualitative measure was used in order to stay away from the asymptotic regime of training, where convergence does continue to improve but the training time grows exponentially. It should be noted that it is possible that further improvements to convergence may be achieved with increased numbers of training steps.

Due to the stochastic nature of the training process, the loss value for each training step can vary. In order to mitigate this, particularly in the case of the variable boundary condition PINNs where this effect is most pronounced, the rolling mean loss was used instead of the instantaneous values. This was done by taking the mean of the most recent 1000 training step loss values. Two different conditions are used in Table \ref{training_and_results} for cessation of training, fixed rolling mean loss and fixed training steps. Results that have been trained to $1\times10^6$ training steps use the latter method, and are present only to demonstrate if any significant improvement to convergence can be achieved with continued training.

\begin{table*}
    \centering
    \begin{tabular}{ |p{2.8cm}|p{2cm}|p{1cm}|p{1.5cm}|p{1.5cm}|p{1.5cm}|p{2.3cm}|p{2.7cm}|  }
    \hline
    \textbf{Result} & \textbf{Trainable parameters} & $\boldsymbol{\sigma_\textbf{train}}$ & \textbf{Rolling Mean Loss} & \textbf{Training Steps} & \textbf{Training Time (s)} & \textbf{Mean PINN Evaluation Time ($\times10^{-4}$s)} & \textbf{Mean Numerical Evaluation Time ($\times10^{-3}$s)} \\
    \hline
    \textbf{A (conventional)}    & 4,674  & 0.0 & $1.0\times10^{-3}$ & $5.11\times10^5$ & 1,639  & $1.19$ & $1.50$\\
    \textbf{B (conventional)}    & 4,674  & 0.0 & $1.5\times10^{-2}$ & $1.58\times10^5$ & 497   & $1.26$ & $1.51$\\
    \textbf{C (conventional)}    & 4,674  & 0.1 & $1.8\times10^{-2}$ & $1.00\times10^6$ & 3,163  & $1.19$ & $1.61$\\
    \textbf{D (conventional)}    & 4,674  & 1.0 & $1.0\times10^{0}$  & $1.00\times10^6$ & 3,103  & $1.14$ & $1.55$\\
    \textbf{E (Plane-Wave)}      & 27,586 & 0.0 & $1.0\times10^{-3}$ & $3.17\times10^4$ & 62    & $1.43$ & $1.60$\\
    \textbf{F (Plane-Wave)}      & 27,586 & 1.0 & $2.0\times10^{-2}$ & $1.40\times10^5$ & 277   & $1.27$ & $1.64$\\
    \textbf{G (Plane-Wave)}      & 27,586 & 1.0 & $1.7\times10^{-2}$ & $1.00\times10^6$ & 1,942  & $1.27$ & $1.58$\\
    \textbf{H (conventional)}    & 25,474 & 0.0 & $1.0\times10^{-3}$ & $1.52\times10^5$ & 400   & $1.18$ & $1.55$\\
    \textbf{I (conventional)}    & 25,474 & 0.1 & $1.7\times10^{-2}$ & $1.00\times10^6$ & 2,625  & $1.03$ & $1.63$\\
    \textbf{J (conventional)}    & 25,474 & 1.0 & $8.0\times10^{-1}$ & $1.00\times10^6$ & 2,613  & $1.01$ & $1.57$\\
    \hline
    \end{tabular}
    \caption{Training and network performance for each result shown in this paper. $\sigma_\text{train}$ is the width of the normal distribution used for sampling the random boundary conditions during training. At regular intervals of $1\times10^4$ training steps, both the PINN and numerical solutions were sampled over the evaluation time domain. The mean PINN and numerical evaluation times were obtained from all of the evaluation intervals during the whole training session. Note that the results for the last three entries are in Appendix A.}
    \label{training_and_results}
\end{table*}

\subsection{Conventional PINN}
\label{conventional_PINN_results}
The first set of results is for the aforementioned toy system of two coupled oscillators, solved using a conventional PINN, for both fixed and variable training boundary conditions. The general form of the architecture used for these PINNs was discussed in \secref{sec:conventional_activation_functions}. The results in the main body of this paper used five hidden layers, comprised of 64 neurons each, whilst Appendix \ref{Append_A} used one hidden layer composed of 64 neurons, followed by three hidden layers composed of 128 neurons.

The first of these two configurations gives rise to a model with 4,674 trainable parameters (as reported by PyTorch), which was chosen since adding significant numbers of parameters did not meaningfully influence the convergence of the model. Whilst this type of PINN is the most flexible (principally owing to it being the most generic and unspecialised), it is usually very slow to converge, resulting in increased GPU memory and compute usage.

\begin{figure}
	\centering
	\includegraphics[width=\linewidth]{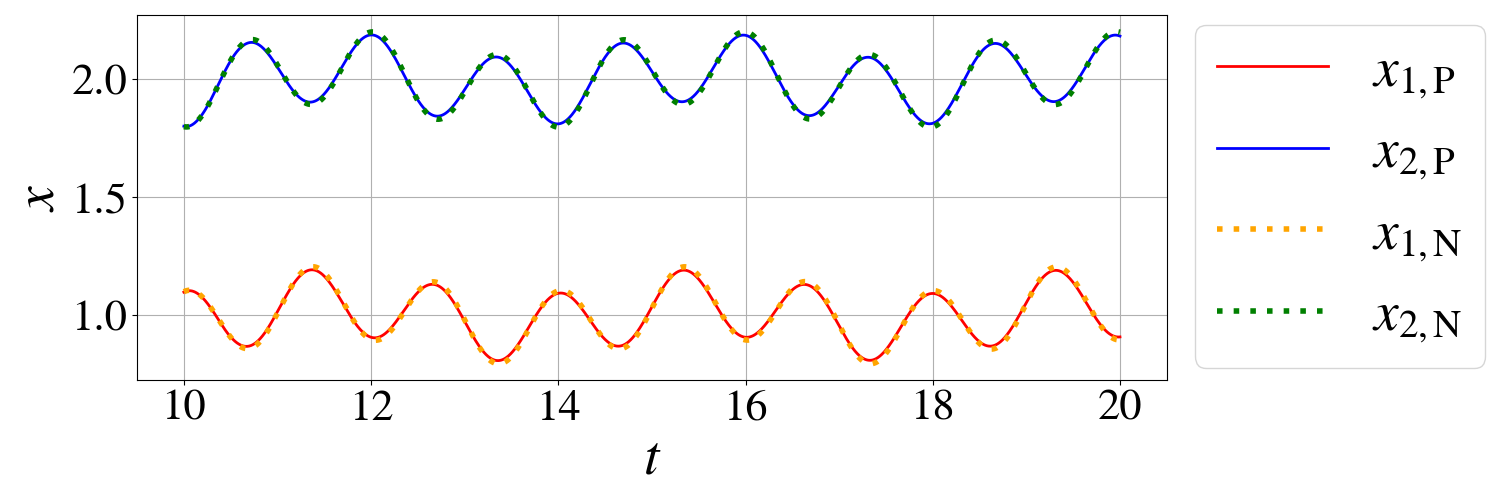}
	\caption{\textbf{Result A}. Solution to a system of two coupled oscillators using a conventional PINN, trained using fixed boundary conditions. $x_{\{1,2\},P}$ and $x_{\{1,2\},N}$ denote the PINN and numerical solutions for both coordinates $x_1$ and $x_2$ respectively.}
	\label{fig:fig_result_A_eval}
\end{figure}

\begin{figure}
	\centering
	\includegraphics[width=\linewidth]{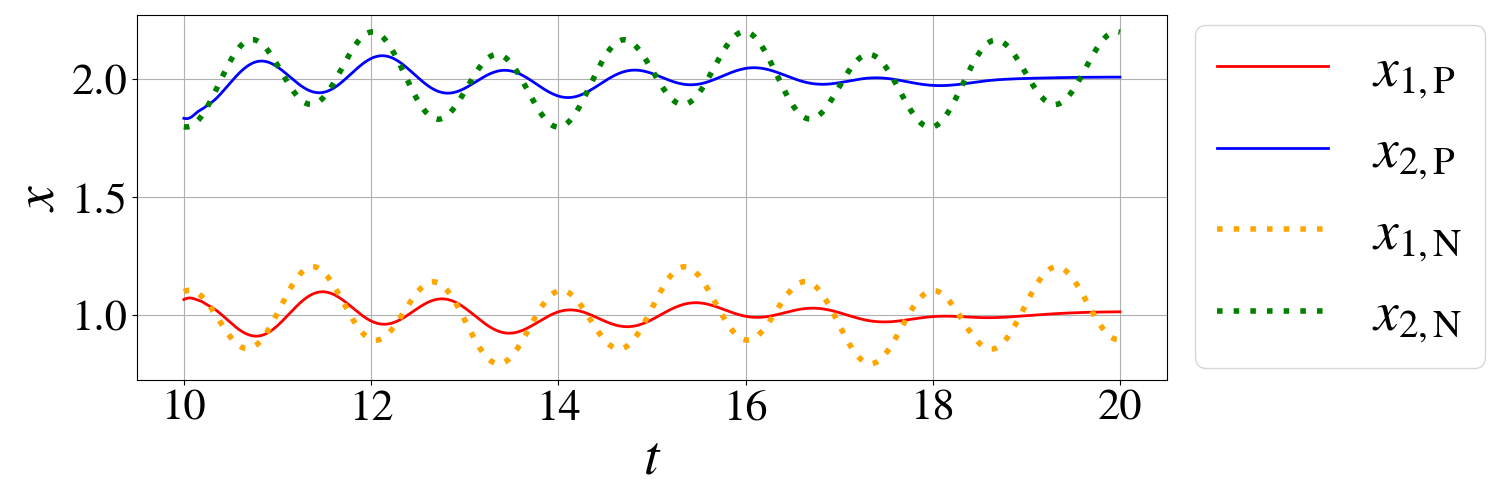}
	\caption{\textbf{Result B}. Solution to a system of two coupled oscillators using a conventional PINN, trained using fixed boundary conditions. $x_{\{1,2\},P}$ and $x_{\{1,2\},N}$ denote the PINN and numerical solutions for both coordinates $x_1$ and $x_2$ respectively. This result was intentionally left in an under-trained state in order to demonstrate how the solution progresses from the boundary conditions to later times $t_a < t \le t_b$ as the training session progresses.}
	\label{fig:fig_result_B_eval}
\end{figure}

The first set of results, shown by \figref{fig:fig_result_A_eval} (\textbf{Result A}) is of a conventional PINN trained to the aforementioned ``reasonable'' convergence on the solution. By cross-referencing these results with Table \ref{training_and_results}, the loss was low but the training time was relatively high, when compared to other results in the table. This result is to demonstrate the application of a conventional PINN to a simple physical system, where only the theoretical model is to be solved and no externally sourced data is used during training. Also of note is that \textbf{Result A} demonstrates that the trained PINN is an order of magnitude faster during evaluation than \texttt{odeint()} is at solving the same problem. However, since this PINN has been trained to solve for exactly one set of unique boundary conditions, the training time must also be included for the solution times to be axiomatically comparable. Any further solutions would require re-training the PINN, and this means it is six orders of magnitude slower than \texttt{odeint()} according to this metric.
A property that this type of PINN exhibits is that the solution appears to be fitted progressively from the boundary conditions to later times during the training process (as number of training steps increases). \textbf{Result B} (\figref{fig:fig_result_B_eval}) illustrates this phenomenon by stopping the training at a poorer state of convergence. This property may arise from the boundary conditions (Dirichlet for the positions and Neumann for the velocities) being the most trivial part of the loss function to fit (the solution simply needs to take the values specified at timestep $t_0$). One could imagine that the boundary loss terms yield a loss landscape that has steep gradients with a single locus around boundary condition optimisation, since they can only take one value each. The same cannot be said for solving the differential equation losses (given by equations \eqref{eq:COL1} and \eqref{eq:COL2}) at later time steps, $t_0 < t \le t_b$, where a continuous distribution of boundary conditions leads to a continuous distribution of unique solutions, and the loss terms yield a loss landscape that is significantly more complicated.
It might seem that for \textbf{Result A} and \textbf{Result B}, the PINN has a rather simple task, since there is only one viable solution that satisfies the single set of boundary conditions and equations of motion. However, in reality the loss landscape formed by the sum of the loss terms \eqref{eq:loss} appears to exhibit steep gradients around the trivial or mean solutions, i.e, where the oscillators are in their equilibrium positions (as discussed above). This is often quite apparent during the training of conventional PINNs, where the solutions stay near the equilibrium positions until the network has been trained sufficiently for the boundary conditions to propagate the unique solution through to these later times. This property of PINN solutions can be controlled by increasing the weighting of the boundary condition loss terms, through one of the $\Lambda$ hyperparameters that scales the relevant terms in the loss function. As with the other results in this paper, \textbf{Result B} may be cross-referenced with Table \ref{training_and_results} for more information on the training and evaluation.

\begin{figure}
	\begin{subfigure}[t]{1\linewidth}
		\includegraphics[width=1\linewidth]{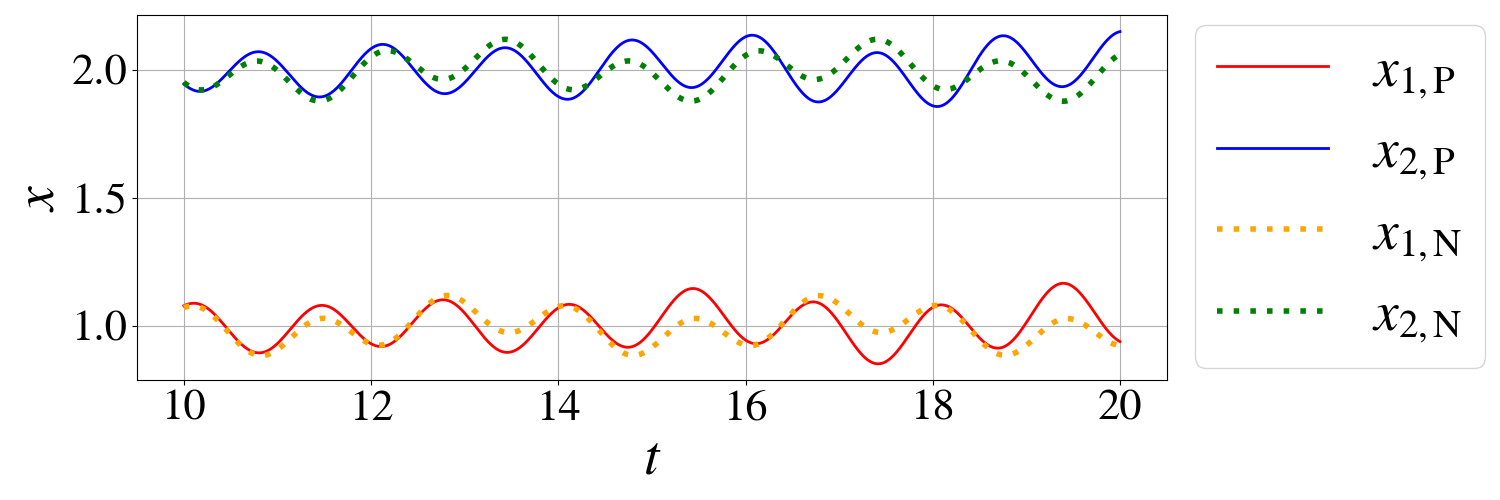}\hfill\centering
		\captionsetup{width = 0.8\linewidth}
		\caption{Training position-time domains and boundary conditions.}
		\vspace{0.5cm}
		\label{fig:fig_result_C_train}
	\end{subfigure}
	\begin{subfigure}[t]{1\linewidth}
		\includegraphics[width=1\linewidth]{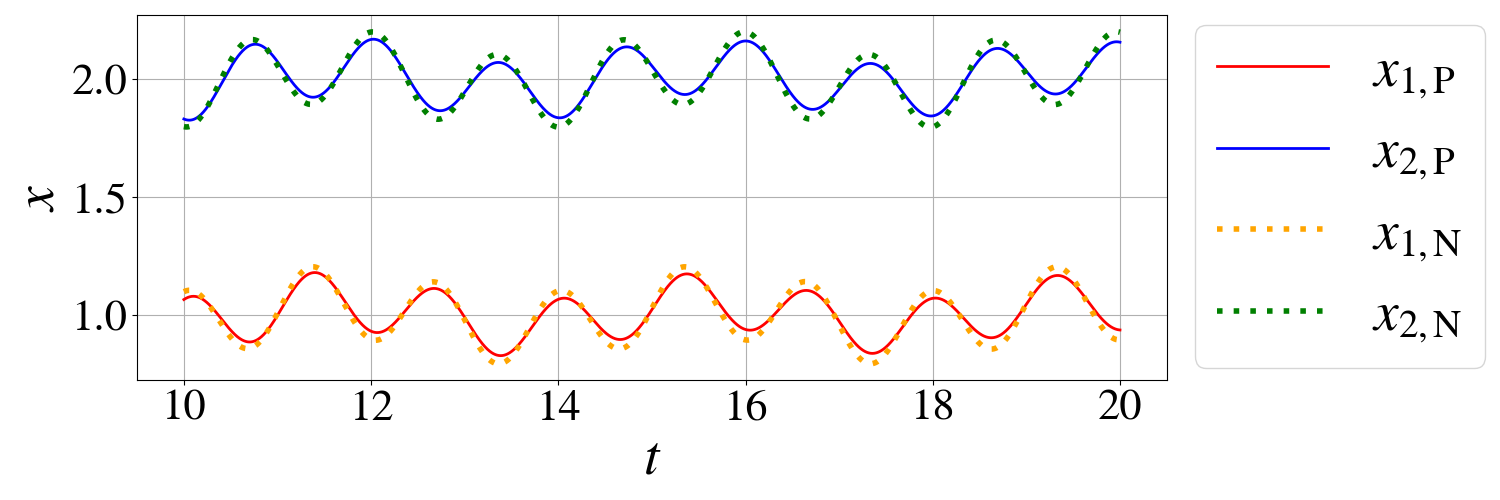}\centering
		\captionsetup{width = 0.8\linewidth}
		\caption{Evaluation position-time domains and boundary conditions.}
		\vspace{0.5cm}
		\label{fig:fig_result_C_eval}
	\end{subfigure}
	\caption{\textbf{Result C}. Solution to a system of two coupled oscillators using a conventional PINN, trained on variable boundary conditions, where $\sigma_\text{train}=0.1$ was used during training. $x_{\{1,2\},P}$ and $x_{\{1,2\},N}$ denote the PINN and numerical solutions for both coordinates $x_1$ and $x_2$ respectively.}
		\label{fig:fig_result_C}
\end{figure}

\begin{figure}
	\begin{subfigure}[t]{1\linewidth}
		\includegraphics[width=1\linewidth]{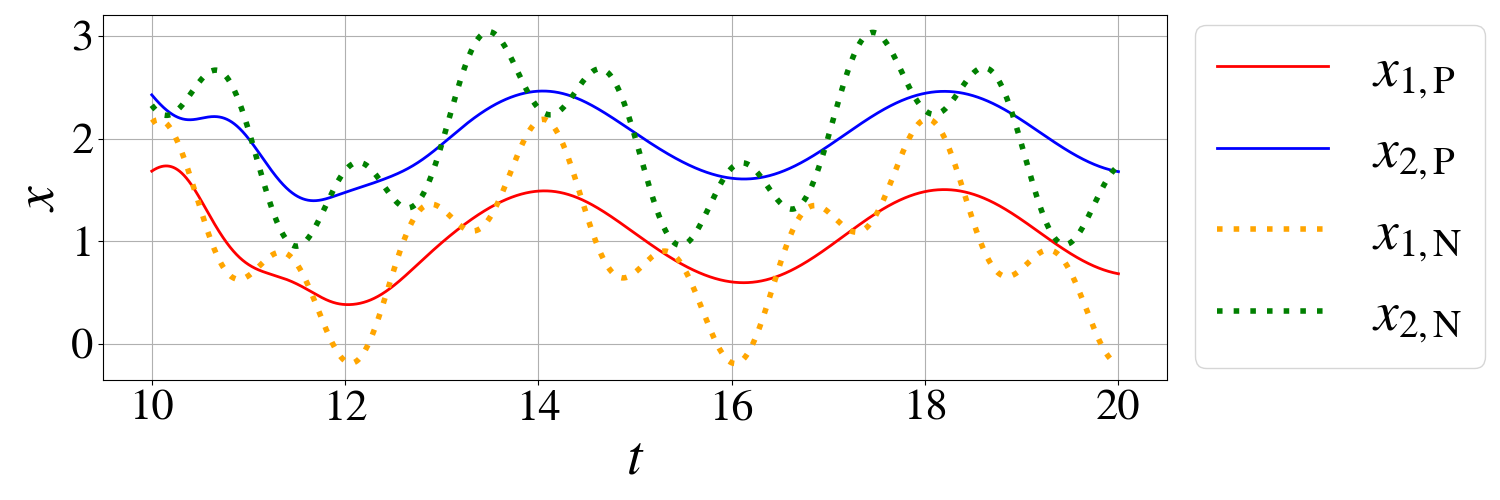}\hfill\centering
		\captionsetup{width = 0.8\linewidth}
		\caption{Training position-time domains and boundary conditions.}
		\vspace{0.5cm}
		\label{fig:fig_result_D_train}
	\end{subfigure}
	\begin{subfigure}[t]{1\linewidth}
		\includegraphics[width=1\linewidth]{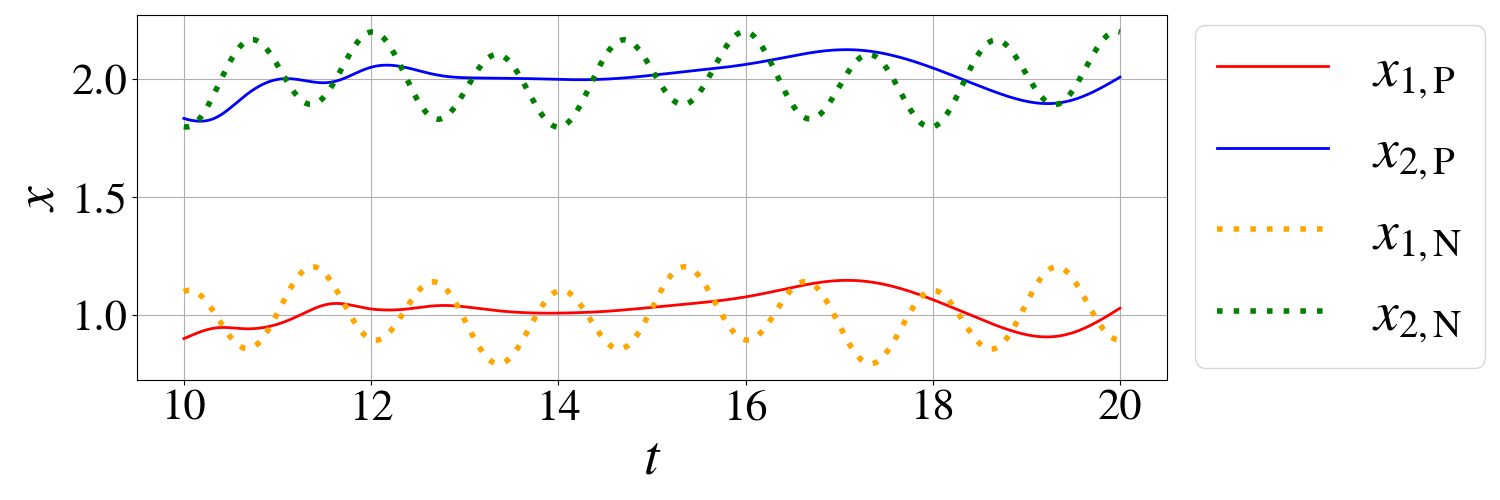}\centering
		\captionsetup{width = 0.8\linewidth}
		\caption{Evaluation position-time domains and boundary conditions.}
		\vspace{0.5cm}
		\label{fig:fig_result_D_eval}
	\end{subfigure}
	\caption{\textbf{Result D}. Solution to a system of two coupled oscillators using a conventional PINN, trained on variable boundary conditions, where $\sigma_\text{train}=1$ was used during training. $x_{\{1,2\},P}$ and $x_{\{1,2\},N}$ denote the PINN and numerical solutions for both coordinates $x_1$ and $x_2$ respectively.}
		\label{fig:fig_result_D}
\end{figure}

Conventional PINNs (at least the ones featured in this paper) do not perform well when an attempt is made to decouple the boundary conditions from the solutions. \textbf{Result C} and \textbf{Result D} shown by figures \figref{fig:fig_result_C} and \figref{fig:fig_result_D} respectively demonstrate this for two different normal distributions (with $\sigma_\text{train}=0.1$ and $\sigma_\text{train}=1$ respectively) used for random sampling of boundary conditions during training. It should be noted that the format of these results differs from the earlier ones, since a sample of the training boundary conditions will, in all probability, be different from the fixed evaluation boundary conditions. To that end, two plots are shown per result. The first is to show the PINN solution for the set of randomly sampled training boundary conditions used for the training step indicated in Table \ref{training_and_results}, and the second showing the PINN's solution for the fixed set of boundary conditions used for evaluation. As discussed in \secref{sec:PINNtraining}, the evaluation boundary conditions remain constant, and take the mean values of the normal distributions used during training. \figref{fig:fig_result_C} (\textbf{Result C}) demonstrates a conventional PINN for which each of the training boundary conditions were sampled from normal distributions with $\sigma_\text{train}=0.1$.
The PINN does begin to converge but the training domain solutions in particular are a qualitatively poor fit to the numerical ones. This remains the case even after the large number of training steps used to generate these results ($1\times10^6$). It may be possible to improve this convergence by using a different network architecture and/or more training steps, but such optimisations are not the focus of this paper.

\textbf{Result D} (shown by \figref{fig:fig_result_D}) demonstrates that increasing the deviation of boundary conditions from the mean values that are used for evaluation only serves to reduce the convergence of the conventional PINN even further from the previous set of results (\textbf{Result C}). In this case, $\sigma_\text{train}=1$ for training and the resulting PINN solutions fail to converge on the numerical ones. Instead, the PINN converges on a time-dependent rolling mean of the equilibrium positions of each oscillator in the training domain (\figref{fig:fig_result_D_train}), and no meaningful conclusions can be drawn from the evaluation domain (\figref{fig:fig_result_D_eval}).
Clearly, the PINN completely fails to converge on the \texttt{odeint()} solution, demonstrating that this simple conventional PINN model cannot easily form a generalised solution that is only made unique during evaluation. Appendix \ref{Append_A} demonstrates that a much larger conventional PINN model, using the same fundamental structure as used here also fails in similar ways.

\subsection{Unique Solution Plane-Wave PINN} \label{sec:unique_PWPINN_results}
As was the case for the conventional PINNs, the general form of the plane-wave PINNs used in this paper was discussed in \secref{sec:plane_wave_PINN_theory}, but the number of neurons in the hidden layers was left undefined. The following results use two $\tanh$-activated hidden layers with 32 and 128 neurons respectively, followed by a sigmoid-activated hidden layer of 100 neurons and then a plane-wave-activated hidden layer of 100 neurons. The 100 neurons in the plane-wave-activated layer are assigned angular frequencies $0 \le \omega \le 5$ with an interval of 0.05 through the plane-wave activation function \eqref{eq:PWActivation}, as discussed in \secref{sec:plane_wave_PINN_theory}. This gives rise to a model with 27,586 trainable parameters (as reported by PyTorch). This is significantly more than in the conventional model featured in \secref{conventional_PINN_results} at 4,674, but unlike the conventional model, more network trainable parameters were required to lead to a network with good performance. It might seem that this could be a contributing factor for why the following results are qualitatively better and converge faster than the previous sets, but we go on to demonstrate in Appendix \ref{Append_A} that an equivalently scaled conventional PINN using the same fully connected architecture utilised in \secref{conventional_PINN_results} does not lead to better convergence or performance. The operation of the conventional and plane-wave PINNs differs significantly, and thus we have not attempted to keep the number of trainable parameters the same. Instead we have attempted to provide examples of network architectures that produce good results, whilst using as few trainable parameters as reasonably possible. However, it should be reiterated that we found that increasing the number of trainable parameters in the conventional PINNs did not meaningfully improve their performance. It should also be noted that we do not mean to say that any of the networks featured here could not be improved upon, as they almost certainly could be. Instead, it is our task in this paper to demonstrate that generalised solutions are not learned well by the conventional PINNs but are by the plane-wave PINNs with relative ease. As stated above, please refer to Appendix \ref{Append_A} for the results of a conventional PINN that has an equivalent number of trainable parameters as the plane-wave PINNs featured in this section.

\begin{figure}
	\centering
	\includegraphics[width=\linewidth]{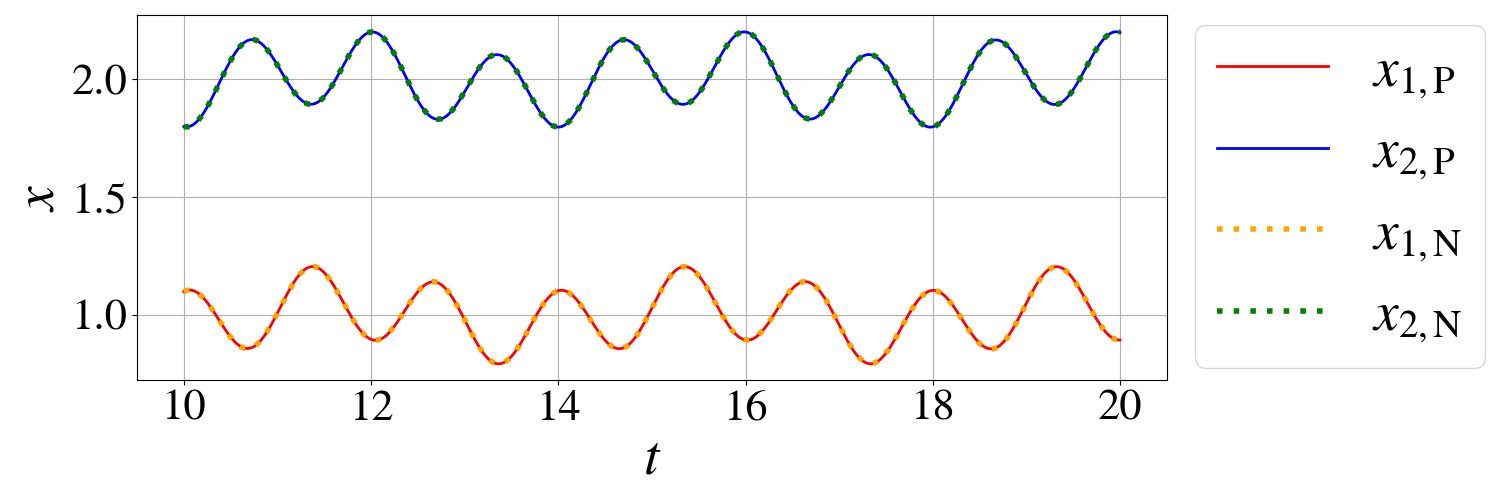}
	\caption{\textbf{Result E}. Solution to a system of two coupled oscillators using unique-solution plane-wave PINNs. $x_{\{1,2\},P}$ and $x_{\{1,2\},N}$ denote the PINN and numerical solutions for both coordinates $x_1$ and $x_2$ respectively.}
	\label{fig:fig_result_E_eval}
\end{figure}

To draw parallels with the conventional PINN solution given by \textbf{Result A} in \secref{conventional_PINN_results}, the first result for the plane-wave PINNs (\textbf{Result E}) shows the solution to the same toy system of coupled oscillators used throughout this paper, utilising plane-wave PINNs that are trained on a fixed set of boundary conditions. This method of training results in the plane-wave PINN learning a single, unique solution to the system. The results for this are shown by \figref{fig:fig_result_E_eval}, depicting \textbf{Result E}.
By referencing Table \ref{training_and_results} it is immediately obvious that there is a drastic reduction in the required number of training steps when compared to the conventional PINN \textbf{Result A}, and a corresponding reduction in training time for results that exhibit the same qualitative convergence on the \texttt{odeint()} solutions and loss. This should come as no surprise since the solution is now expressed in a more natural basis, meaning the layers of the PINN preceding the plane-wave layer are purely tasked with finding the scaling factor for each plane-wave. By approaching the problem in this way, the plane-wave PINN achieves a solution with an equivalent loss to the conventional PINN, but in $1/26$ of the training time. A key point to note about this network is that the set of boundary conditions is the only information that all but the last two layers of the network is dependent on. The time domain is only introduced in the penultimate (plane-wave) layer, before final summation in the last layer, such that the output is a sum of individually scaled plane waves. The task for the first layers of the network is to find the multiplicative prefactor for each plane-wave, as discussed in \secref{sec:plane_wave_PINN_theory}.

\subsection{Generalised Solution Plane-Wave PINN} \label{sec:GSPWPINNs}
\begin{figure}
	\begin{subfigure}[t]{1\linewidth}
		\includegraphics[width=1\linewidth]{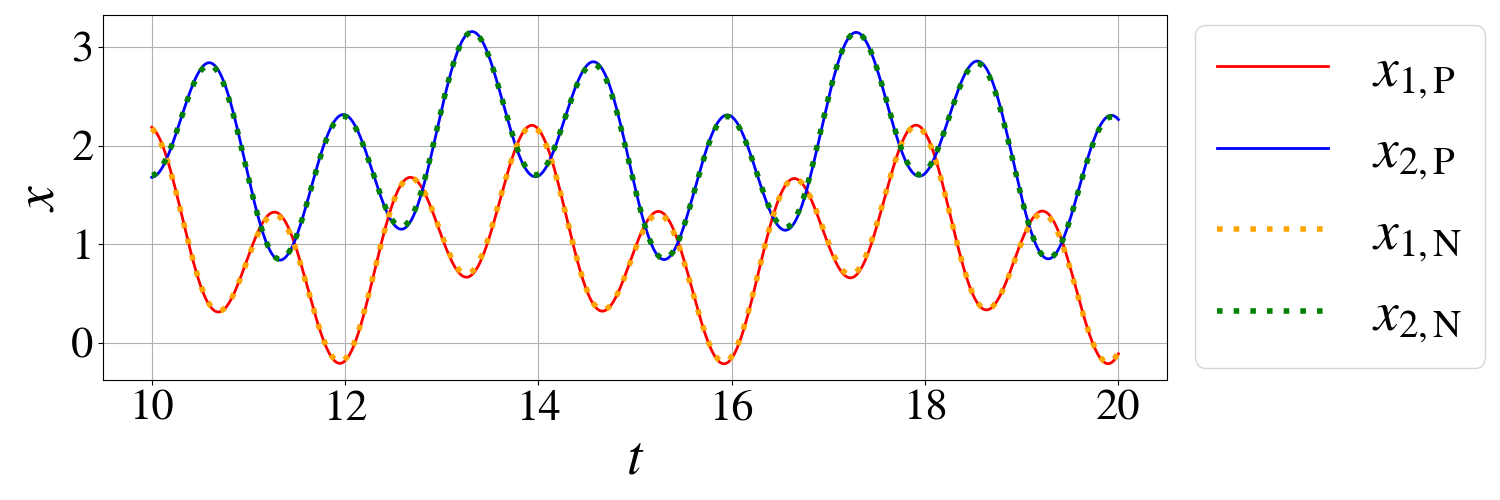}\hfill\centering
		\captionsetup{width = 0.8\linewidth}
		\caption{Training position-time domains and boundary conditions.}
		\vspace{0.5cm}
		\label{fig:fig_result_F_train}
	\end{subfigure}
	\begin{subfigure}[t]{1\linewidth}
		\includegraphics[width=1\linewidth]{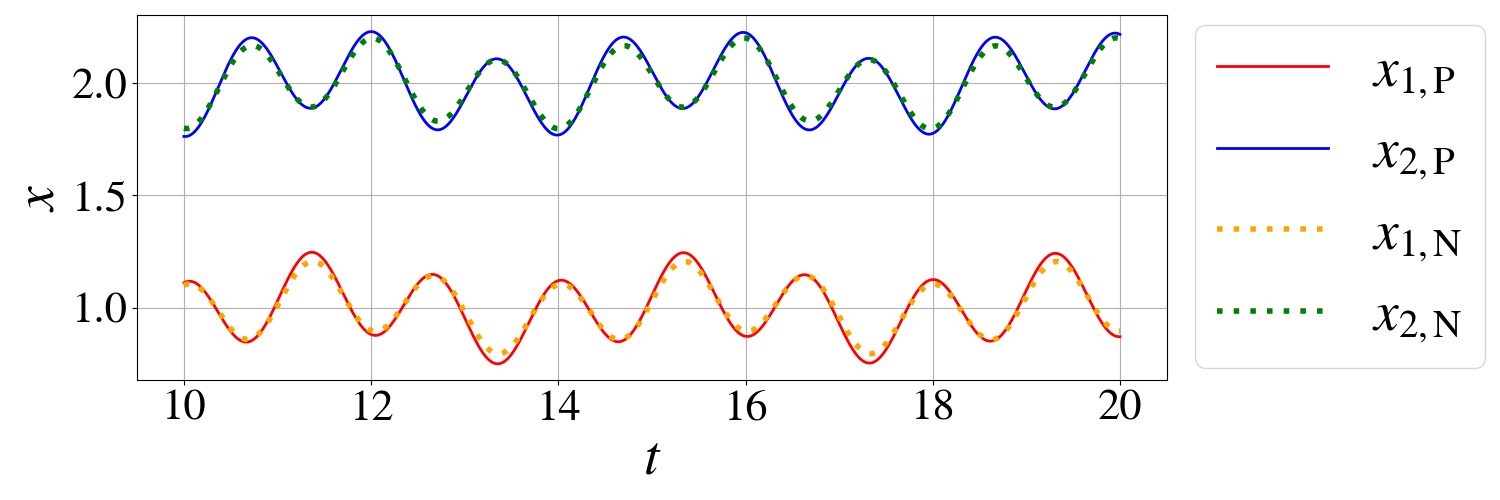}\centering
		\captionsetup{width = 0.8\linewidth}
		\caption{Evaluation position-time domains and boundary conditions.}
		\vspace{0.5cm}
		\label{fig:fig_result_F_eval}
	\end{subfigure}
	\caption{\textbf{Result F}. Solution to a system of two coupled oscillators using generalised-solution plane-wave PINNs, trained on variable boundary conditions, where $\sigma_\text{train}=1$ was used during training. $x_{\{1,2\},P}$ and $x_{\{1,2\},N}$ denote the PINN and numerical solutions for both coordinates $x_1$ and $x_2$ respectively.}
		\label{fig:fig_result_F}
\end{figure}

As discussed in \secref{sec:decoupled_bc_theory}, the PINNs used in \secref{sec:unique_PWPINN_results} are adapted to train on decoupled boundary conditions by supplying the input layer with random samples taken from normal distributions for which $\sigma_\text{train} > 0$. The same samples supplied to the input of the PINN are also used in the boundary condition loss terms, for each training step respectively. Again, these PINNs are tasked with solving the same toy system of two coupled oscillators used throughout this paper. \figref{fig:fig_result_F} depicts \textbf{Result F}, the solutions obtained using the generalised-solution plane-wave PINN, which was trained using normal distributions for the boundary conditions for which $\sigma_\text{train}=1$.
Whilst this network did take more than four times longer to train than the unique solution plane-wave PINN (\textbf{Result E}) (see Table \ref{training_and_results}), the trained PINN generates good solutions with a relatively minor penalty to the qualitative fit to the numerical solutions. In addition, \textbf{Result F} was trained in 277s, versus the 3103s for \textbf{Result D} (the conventional PINN counterpart to this simulation). The plane-wave PINN used to generate \textbf{Result F} would appear to have learned a general solution, which is then made unique only during evaluation rather than training. To investigate this, a range of solutions was generated using different boundary conditions, and the difference between these and the equivalent numerical solutions was plotted as a surface.

\begin{figure}
	\centering
	\includegraphics[width=75mm]{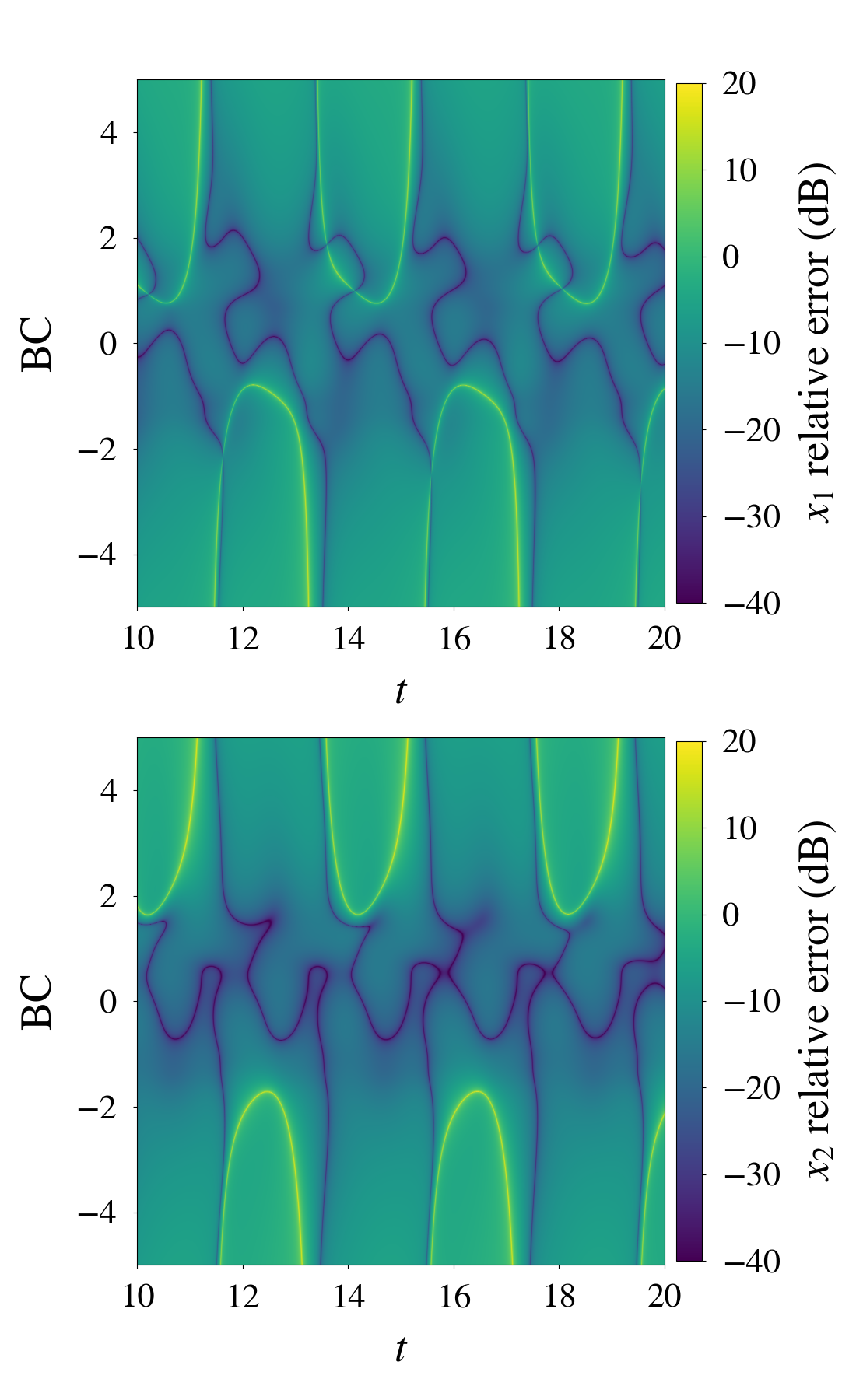}
	\caption{\textbf{Result F}. Difference surfaces between the PINN and numerical solution spectra for both coordinates where the spectrum is formed by a range of boundary conditions and associated position-time solutions. This PINN was trained to a loss of $2\times10^{-2}$.}
	\label{fig:fig_result_F_diff_surf}
\end{figure}

\begin{figure}
	\centering
	\includegraphics[width=75mm]{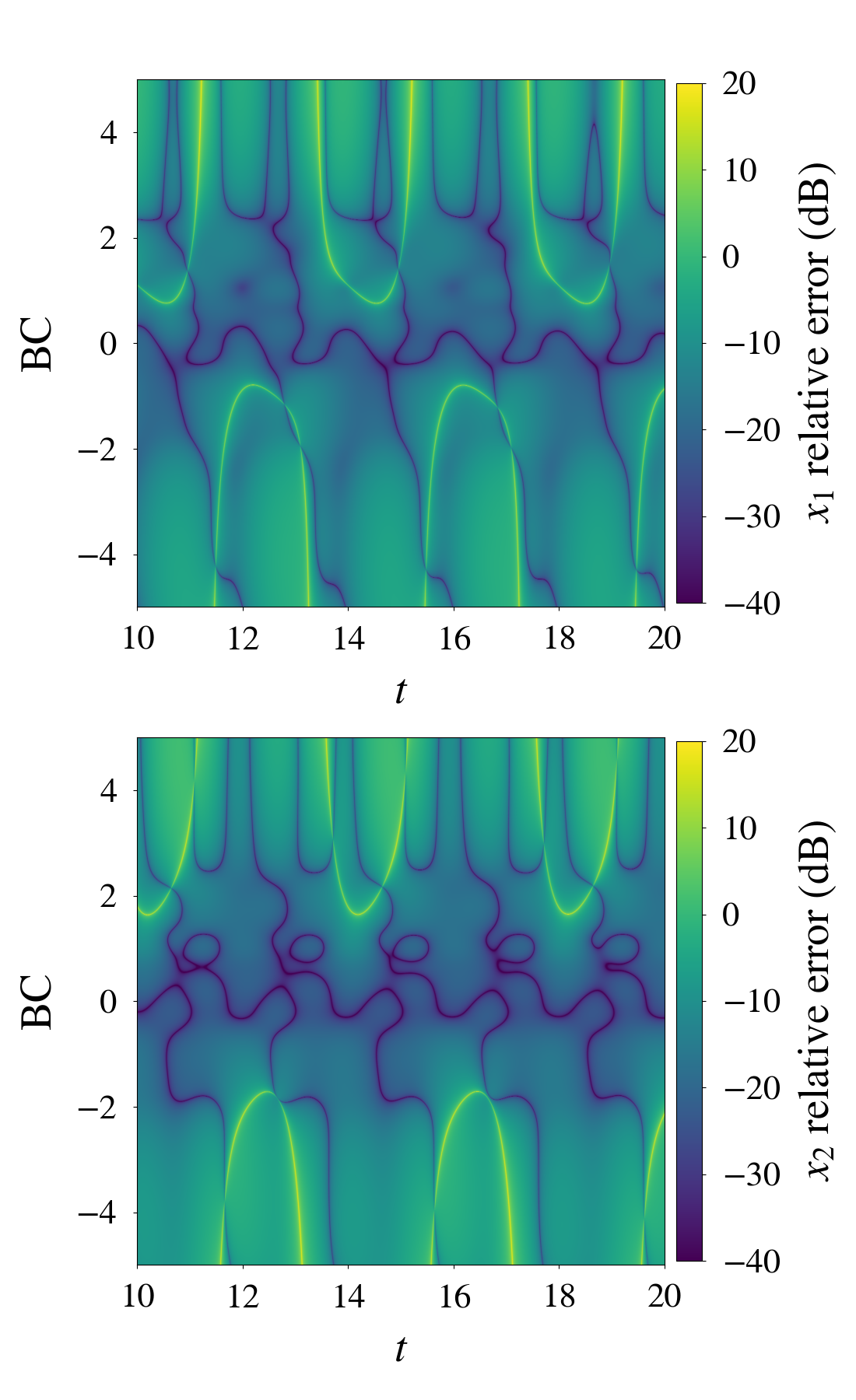}
	\caption{\textbf{Result G}. Difference surfaces between the PINN and numerical solution spectra for both coordinates where the spectrum is formed by a range of boundary conditions and associated position-time solutions. This PINN was trained to $1\times10^6$ training steps.}
	\label{fig:fig_result_G_diff_surf}
\end{figure}

In order to get representative (but not exhaustive) samples of how the network performs in the four-dimensional volume described by the variation of all four boundary conditions ($\delta x_1(t_0)$, $\delta \dot{x}_1(t_0)$, $\delta x_2(t_0)$, $\delta \dot{x}_2(t_0)$) each one was incremented simultaneously, such that the path described between $x_1(t_0)$, $\dot{x}_1(t_0)$, $x_2(t_0)$, $\dot{x}_2(t_0)$ and $x_1(t_b), \dot{x}_1(t_b), x_2(t_b), \dot{x}_2(t_b)$ is a straight line in the four-dimensional volume. \figref{fig:fig_result_F_diff_surf} depicts \textbf{Result F}, showing the difference surface constructed by the subtracting the PINN solution surface from the numerical one for the same range of boundary conditions. The range of boundary conditions over which the trained PINN was evaluated is five times $\sigma_\text{train}$ for each boundary condition in the set, such that the PINN's behavior may be examined for conditions that were significantly outside those that the training session is likely to have exposed it to. Both \figref{fig:fig_result_F_diff_surf} and \figref{fig:fig_result_G_diff_surf} display the relative error in decibels, given by
\begin{equation} \label{eq:rel_error_dB}
    \text{ relative error} = 10\log_{10} \left( \frac{\left| x_{\{1,2\},P} - x_{\{1,2\},N} \right|}{\left| x_{\{1,2\},N} \right|} \right)
\end{equation}
where the $P$ and $N$ subscripts denote the PINN and numerical solutions respectively for coordinates $x_1$ and $x_2$.
The difference surfaces are shown by \figref{fig:fig_result_F_diff_surf} and \figref{fig:fig_result_G_diff_surf}, depicting \textbf{Result F} and \textbf{Result G} respectively and demonstrate that the PINN is capable of extrapolating to a degree outside its training domain of boundary conditions as well as interpolating within it. Note that in order to avoid extremely negative or positive results, the relative error in both of these plots is cut off to keep them in the range of -40dB to +20dB.
It is worth reiterating that due to the way the combination of the boundary conditions is varied, the extreme limits of the solution spectra represent the least probable samples on which the PINN will have been trained (since these will be the combination of the tails of four normal distributions, each with $\sigma_\text{train} = 1$). The blue lines that wind through the plots in \figref{fig:fig_result_F_diff_surf} are regions of the lowest relative error, whilst the yellow lines that bound the periodic intrusions from the lower and upper boundary condition limits show where the solutions begin to fail. Interestingly, there appears to be a skew towards better solutions for the second coordinate $x_2$, as the first demonstrates a smaller area for which the solutions exhibit a reasonable convergence.

\begin{figure}
	\begin{subfigure}[t]{1\linewidth}
		\includegraphics[width=1\linewidth]{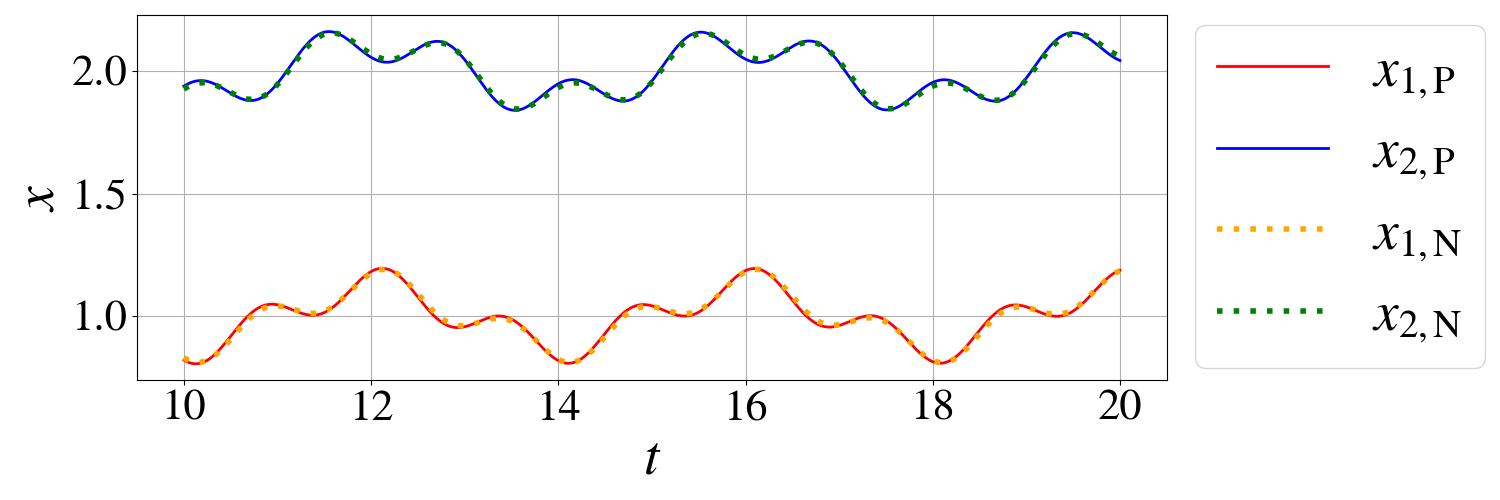}\hfill\centering
		\captionsetup{width = 0.8\linewidth}
		\caption{Training position-time domains and boundary conditions.}
		\vspace{0.5cm}
		\label{fig:fig_result_G_train}
	\end{subfigure}
	\begin{subfigure}[t]{1\linewidth}
		\includegraphics[width=1\linewidth]{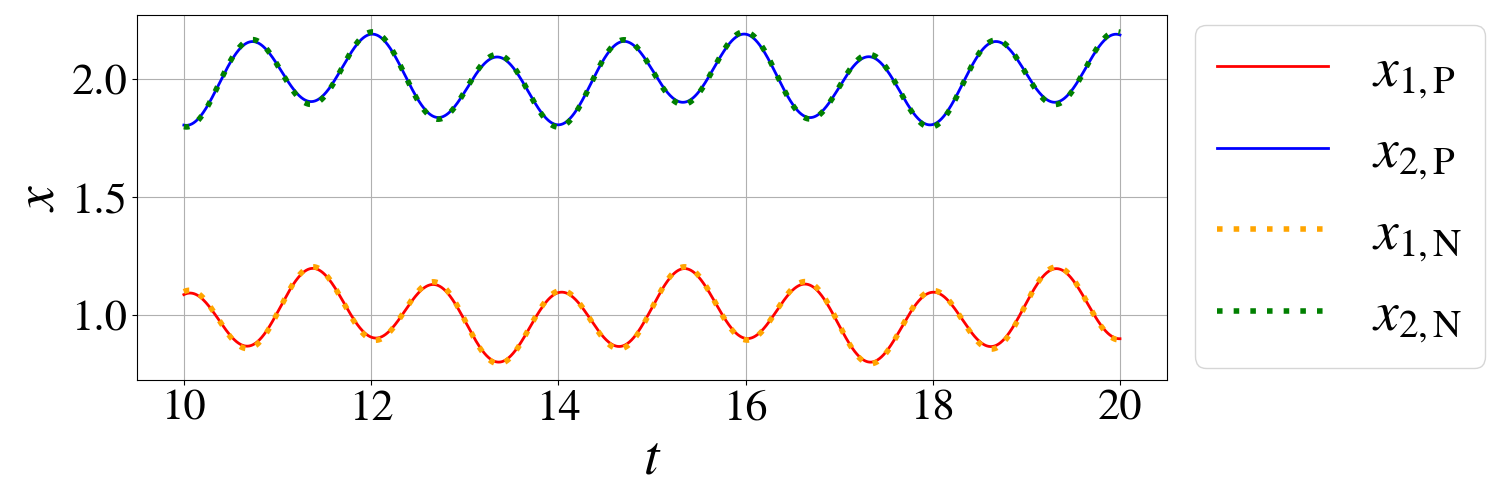}\centering
		\captionsetup{width = 0.8\linewidth}
		\caption{Evaluation position-time domains and boundary conditions.}
		\vspace{0.5cm}
		\label{fig:fig_result_G_eval}
	\end{subfigure}
	\caption{\textbf{Result G}. Solution to a system of two coupled oscillators using generalised-solution plane-wave PINNs, trained on variable boundary conditions, where $\sigma_\text{train}=1$ was used during training. $x_{\{1,2\},P}$ and $x_{\{1,2\},N}$ denote the PINN and numerical solutions for both coordinates $x_1$ and $x_2$ respectively.}
	\label{fig:fig_result_G}
\end{figure}

For completeness, \textbf{Result G} has been included and is shown by \figref{fig:fig_result_G} and \figref{fig:fig_result_G_diff_surf}. In this case the PINN is unaltered from the previous \textbf{Result F} other than continuing the training process to $1\times10^6$ training steps, such that any further improvement to the solution-difference surfaces may be investigated.
Again, the results have been plotted using relative error in decibels, given by \eqref{eq:rel_error_dB}. It can be seen that there is an increased density of blue lines denoting extremely low error, but the solution-difference surfaces depicted by \figref{fig:fig_result_G_diff_surf} shows little to no improvement to the area of low error for either coordinate. This is both encouraging and troubling, since it means the PINN does not meaningfully benefit from further training. This is good from a computational resources point of view, since the PINN rapidly converges on the solution. However, it does also indicate that more research/experimentation is required in order to improve convergence further -- as continued training does not meaningfully do so.

\section{Conclusions\protect\\}\label{conclusions}
PINNs are a recent advancement in solving physical problems. By using the theoretical model expressed by governing equation(s) alone in the loss function, they can be used as an alternative to existing numerical solvers. This represents a use case for artificial intelligence outside the traditional mapping of $X$ and $Y$ training datasets. To date however, the approaches used for ``pure-physics'' PINNs and their semantic equivalents has meant that they very often only produce unique solutions, since they are trained for specific boundary conditions. This is a difficult problem to solve using conventional activation functions alone, resulting in a long and computationally expensive training process even when generating a single unique solution. In practice, this means that in a lot of cases when PINNs are used in this way they represent a very slow alternative to popular numerical methods. Philosophically, in mathematical physics we understand that problems are likely to exhibit solutions with properties that can be leveraged to accelerate the finding of them. As has been the subject of this paper, we aimed to make our PINNs reflect this policy. Further work is required to investigate systems that exhibit qualitative shifts in behavior upon altering the boundary conditions, such as bifurcation, in order to determine if the PINNs featured in this work remain viable in these cases too.

In this paper we demonstrate an approach to structuring a PINN for a toy system of two coupled oscillators. We use our understanding of the expected solution (a basis set of plane waves) to inform our choice of activation functions, in order to provoke the networks into solving the system through the use of meaningful decompositions of the aforementioned expected solutions, such as plane waves. Additionally, we introduce the boundary conditions as variables into the system and train the network over a range of randomly selected values. The resulting network learns the general form of the solution, independent of the boundary conditions. This trained PINN can then find a unique solution during evaluation by passing it a set of boundary conditions. The evaluation of the trained PINNs featured in this paper is an order of magnitude faster than \texttt{odeint()} is at solving the same system, and we show its solutions can be quite accurate even when tasked with solving the system for boundary conditions significantly outside those it has been trained over. This approach has the potential to massively expand the utility of PINNs, particularly for systems where a general solution is expected to be evaluated a large number times for different sets of boundary conditions within some range.

\clearpage
\appendix
\section{A Larger Conventional PINN} \label{Append_A}
\begin{figure}[H]
	\centering
	\includegraphics[width=\linewidth]{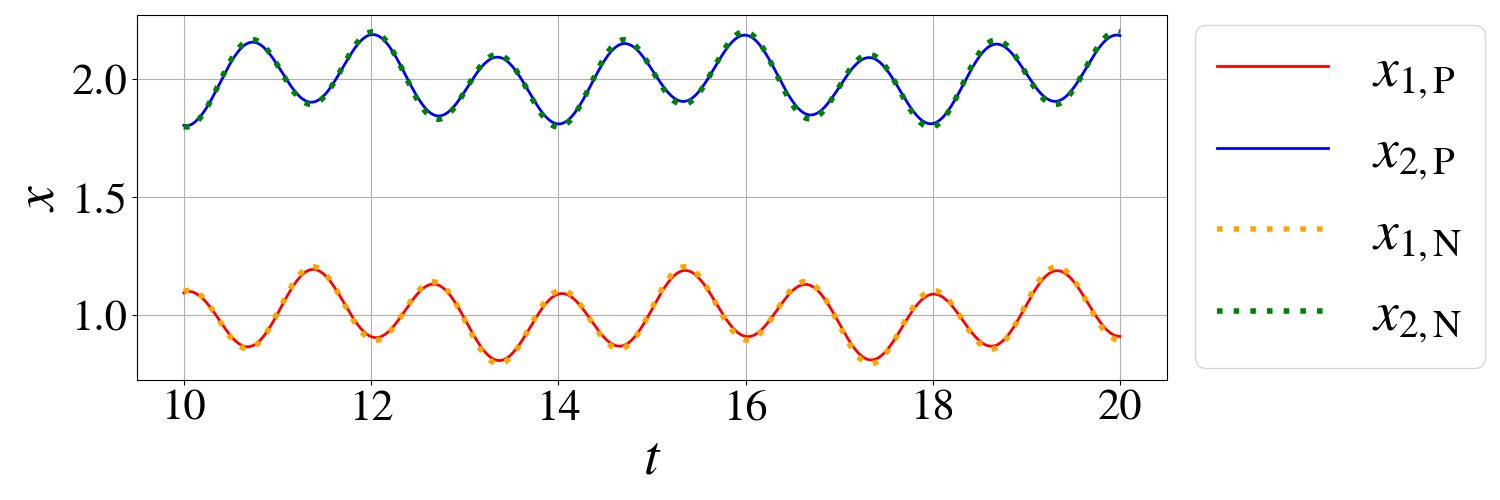}
	\caption{\textbf{Result H}. Solution to a system of two coupled oscillators using a conventional PINN, trained on fixed boundary conditions. $x_{\{1,2\},P}$ and $x_{\{1,2\},N}$ denote the PINN and numerical solutions for both coordinates $x_1$ and $x_2$ respectively. The conventional PINN used for this result has a similar number of trainable parameters as the plane-wave PINNs used in the main body of this paper.}
	\label{fig:fig_result_H_eval}
\end{figure}

As discussed in \secref{sec:unique_PWPINN_results}, the results for a conventional PINN model that has a similar number of trainable parameters as the plane-wave PINNs have been included. It is difficult to obtain a PINN with exactly the same number of trainable parameters, so instead we have chosen to aim for approximately the same number of trainable parameters and to use the same number of hidden layers as was used in the plane-wave PINNs instead. The following results use one hidden layer of 64 neurons, followed by three further hidden layers of 128 neurons, which gives rise to 25,474 trainable parameters, close to the 27,586 parameters in the plane-wave PINN models utilised in the main body of this paper. \figref{fig:fig_result_H_eval} shows \textbf{Result H}, which is the same simulation as \textbf{Result A}, but using the larger conventional network and is trained to the same loss, given in Table \ref{training_and_results}.
No significant qualitative improvement was made to the solutions by increasing the number of network parameters, other than reducing the training time by a factor of four. Whilst this does represent a significant reduction, it is still six times longer than the equivalent plane-wave PINN result, \textbf{Result E}.

\begin{figure}
	\begin{subfigure}[t]{1\linewidth}
		\includegraphics[width=1\linewidth]{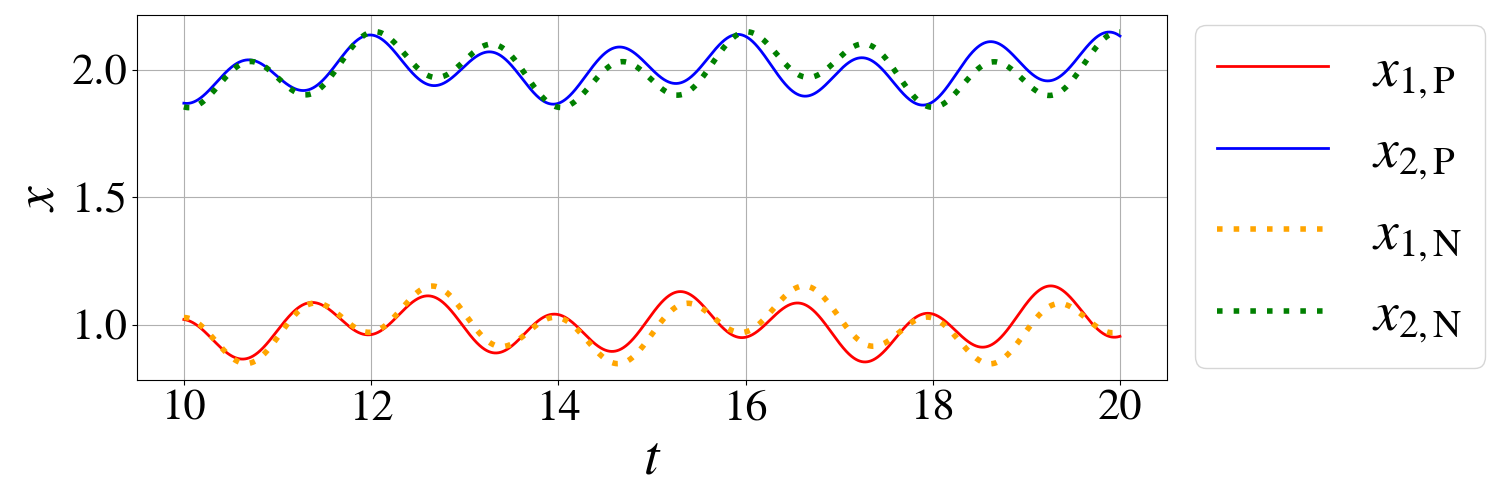}\hfill\centering
		\captionsetup{width = 0.8\linewidth}
		\caption{Training position-time domains and boundary conditions.}
		\vspace{0.5cm}
		\label{fig:fig_result_I_train}
	\end{subfigure}
	\begin{subfigure}[t]{1\linewidth}
		\includegraphics[width=1\linewidth]{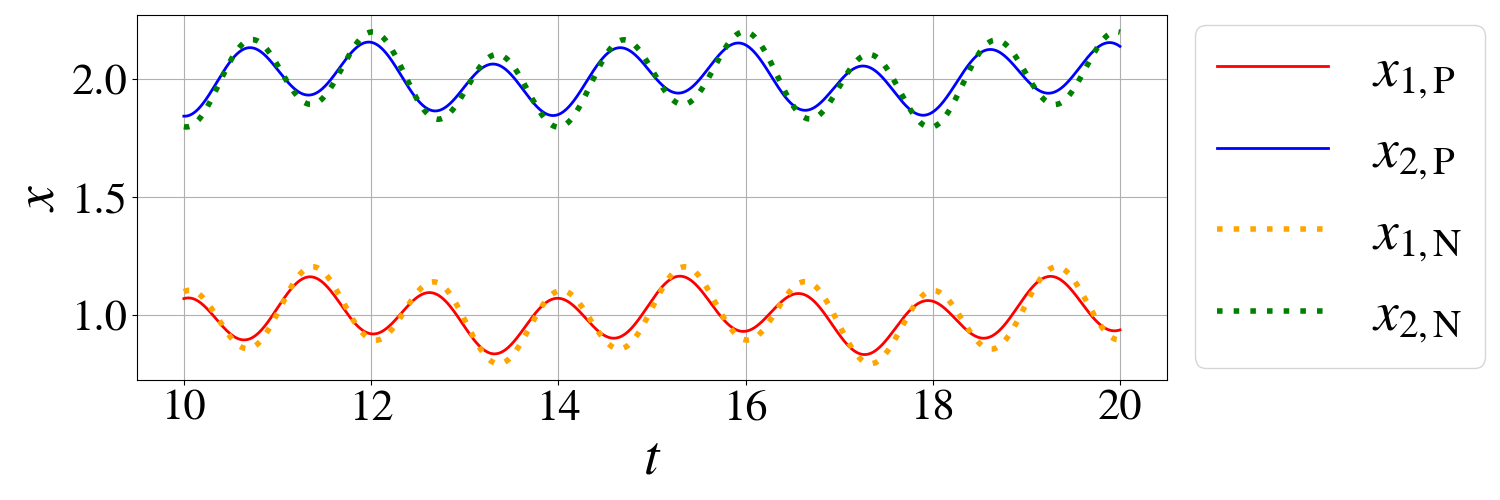}\centering
		\captionsetup{width = 0.8\linewidth}
		\caption{Evaluation position-time domains and boundary conditions.}
		\vspace{0.5cm}
		\label{fig:fig_result_I_eval}
	\end{subfigure}
	\caption{\textbf{Result I}. Solution to a system of two coupled oscillators using a conventional PINN, trained on variable boundary conditions, where $\sigma_\text{train}=0.1$ was used during training. $x_{\{1,2\},P}$ and $x_{\{1,2\},N}$ denote the PINN and numerical solutions for both coordinates $x_1$ and $x_2$ respectively. The conventional PINN used for this result has a similar number of trainable parameters as the plane-wave PINNs.}
	\label{fig:fig_result_I}
\end{figure}

\figref{fig:fig_result_I} shows \textbf{Result I}, which is the same simulation as \textbf{Result C} where $\sigma_\text{train}=0.1$, but using the larger conventional PINN described above. This training session was run to $1\times10^6$ training steps, in order to correspond with \textbf{Result C}.
This result looks qualitatively similar to the corresponding result in the main body of this paper for the smaller conventional network (\textbf{Result C}), but by referencing Table \ref{training_and_results} it can be seen that there is no significant difference in the loss or training times.

\begin{figure}
	\begin{subfigure}[t]{1\linewidth}
		\includegraphics[width=1\linewidth]{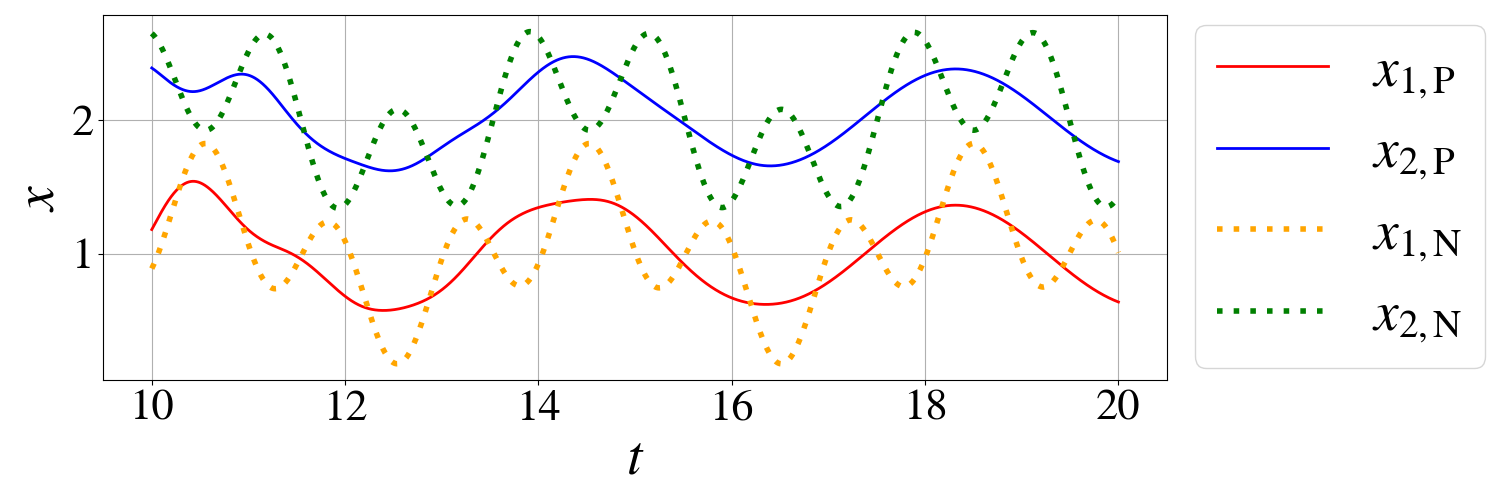}\hfill\centering
		\captionsetup{width = 0.8\linewidth}
		\caption{Training position-time domains and boundary conditions.}
		\vspace{0.5cm}
		\label{fig:fig_result_J_train}
	\end{subfigure}
	\begin{subfigure}[t]{1\linewidth}
		\includegraphics[width=1\linewidth]{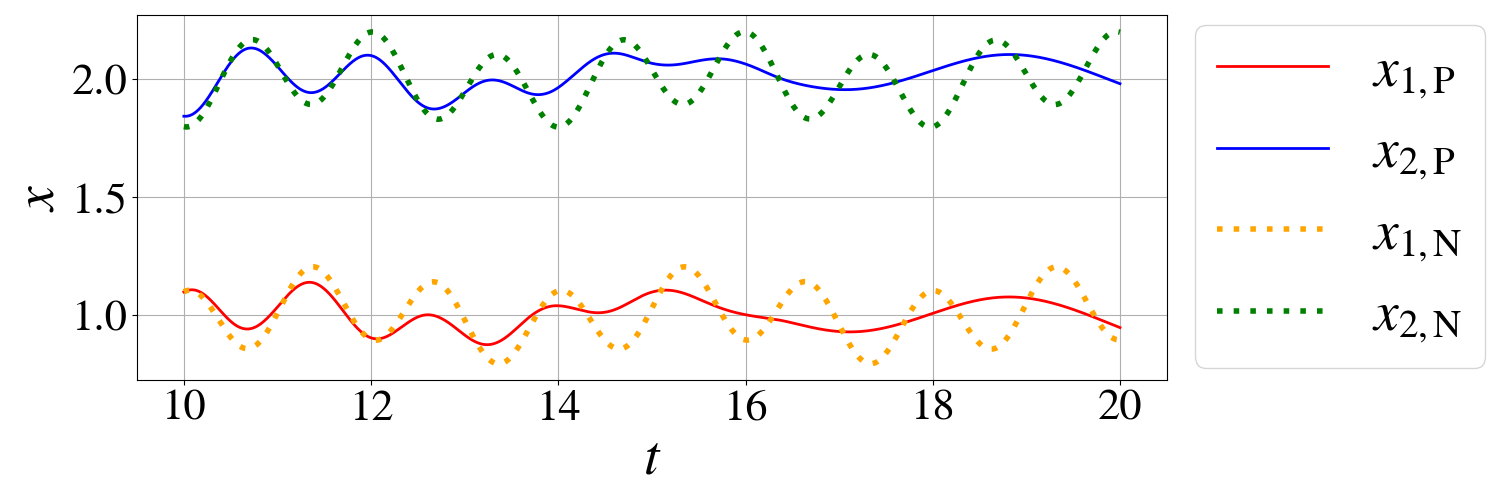}\centering
		\captionsetup{width = 0.8\linewidth}
		\caption{Evaluation position-time domains and boundary conditions.}
		\vspace{0.5cm}
		\label{fig:fig_result_J_eval}
	\end{subfigure}
	\caption{\textbf{Result J}. Solution to a system of two coupled oscillators using a conventional PINN, trained on variable boundary conditions, where $\sigma_\text{train}=1$ was used during training. $x_{\{1,2\},P}$ and $x_{\{1,2\},N}$ denote the PINN and numerical solutions for both coordinates $x_1$ and $x_2$ respectively. The conventional PINN used for this result has a similar number of trainable parameters as the plane-wave PINNs.}
	\label{fig:fig_result_J}
\end{figure}

Finally, \figref{fig:fig_result_J} shows \textbf{Result J}, which is the same simulation as \textbf{Result D} where $\sigma_\text{train}=1$, but using the larger conventional PINN described above. This training session was again run to $1\times10^6$ training steps.
It is difficult to draw any meaningful conclusions from this result, along with its corresponding smaller conventional PINN model (\textbf{Result D}) since they both fail to converge on the numerical solutions completely. As discussed in \secref{conventional_PINN_results}, the solutions appear to be a poor approximation based on a time-dependent rolling mean of the equilibrium positions. This result was only included to demonstrate that the conventional PINNs featured in this work cannot easily generalise when $\sigma_\text{train} > 0$ to the same extent as a plane-wave PINN can, even when the number of network trainable parameters are similar.

\newpage
\bibliography{references}

\end{document}